\documentclass[12pt]{article}
\usepackage[utf8]{inputenc}

\usepackage[letterpaper,
top    = 1.2in,
bottom = 1.2in,
left   = 1.2in,
right  = 1.2in]{geometry}

\usepackage{amsfonts}
\usepackage{amsmath}
\usepackage{amssymb}
\usepackage{bm}
\usepackage[labelfont=bf]{caption}
\usepackage{natbib}
\usepackage{graphicx}
\usepackage{subcaption}
\usepackage[svgnames]{xcolor}
\usepackage{hyperref}
\usepackage{multirow}
\usepackage{enumitem}
\usepackage{setspace}
\usepackage{footmisc}
\usepackage{fancyhdr}

\numberwithin{equation}{section}

\title{Analyzing Stochastic Computer Models: A Review with Opportunities}
\author{
Evan Baker\thanks{Primary and corresponding author: Department of Mathematics, University of Exeter; \href{mailto:e.baker@exeter.ac.uk}{\tt e.baker@exeter.ac.uk}}
\and
Pierre Barbillon\thanks{UMR MIA-Paris, AgroParisTech, INRA, Université Paris-Saclay, 75005, Paris, France}
\and
Arindam Fadikar\thanks{Argonne National Laboratory}
\and
Robert B.~Gramacy\thanks{Department of Statistics, Virginia Tech}
\and
Radu Herbei\thanks{Department of Statistics, The Ohio State University}
\and
David Higdon\footnotemark[4]
\and
Jiangeng Huang\thanks{Department of Statistics, University of California, Santa Cruz}
\and
Leah R.~Johnson\footnotemark[4]
\and
Pulong Ma \thanks{The Statistical and Applied Mathematical Sciences Institute and Duke University}
\and
Anirban Mondal\thanks{Department of Mathematics, Applied Mathematics, and Statistics, Case Western Reserve University}
\and
Bianica Pires\thanks{The MITRE Corporation}
\and
Jerome Sacks\thanks{National Institute of Statistical Sciences}
\and
Vadim Sokolov\thanks{Systems Engineering and Operations Research, George Mason University}
}
\date{}

\begin{document}

\maketitle
\begin{abstract}
In modern science, computer models are often used to understand complex phenomena, and a thriving statistical community has grown around analyzing them. This review aims to bring a spotlight to the growing prevalence of stochastic computer models --- providing a catalogue of statistical methods for practitioners, an introductory view for statisticians (whether familiar with deterministic computer models or not), and an emphasis on open questions of relevance to practitioners and statisticians. Gaussian process surrogate models take center stage in this review, and these, along with several extensions needed for stochastic settings, are explained. The basic issues of designing a stochastic computer experiment and calibrating a stochastic computer model are prominent in the discussion. Instructive examples, with data and code, are used to describe the implementation of, and results from, various methods.
\end{abstract}

{\it Keywords: Computer Model; Gaussian Process; Uncertainty Quantification; Emulator; Computer Experiment; Agent Based Model; Surrogates; Calibration}

\section{Introduction}
\label{sec:intro}
Computer models, also known as simulators, are in use everywhere. These are programs which describe and approximate a process of interest. The code typically takes a set of inputs and produces some output. Stochastic simulators, unlike deterministic ones, can produce different output with the same inputs due to the presence of random elements.\footnote{This terminology can have different meanings and connotations in different fields. In weather modeling, a stochastic simulator might refer specifically to a random weather generator \citep{richardson1981stochastic, peleg2017advanced}. In this work, we use the term to refer to any code that includes pseudo-random deviates in generating output.} Such computer models are in wide use. For example, agent-based models (ABMs) deal with large populations of individuals, where specific actions taken at any one time-step have complexities and uncertainties that do not allow deterministic modeling. ABMs are prevalent \citep{johnson2010implications, johnson2011parameter, ramsey2010management, smieszek2011reconstructing, grimm2006standard} and used to explore complex phenomena in sociology, transportation, ecology, epidemiology, and other phenomena.

The following is a basic model of a stochastic simulator experiment. If the code is run at a (vector) input $x$ producing a (scalar) output $y(x)$, this could be represented as:
\begin{equation}
y(x) = M(x) + v, \ v \sim N(0, \sigma_v^2(x)),
\label{eq:basic}
\end{equation}
where $M(x)$ is the expected value, $E[y(x)]$, of the output. The variability $v$ accounts for the randomness of the stochastic simulator, ultimately caused by pseudo-random number generation within the code. Its variance, $\sigma^2_v$, often depends on $x$, with constant variance subsumed as a special case. For deterministic simulators, $\sigma^2_v = 0$.

Randomness in stochastic simulators invariably requires many simulations thereby limiting the complexity (including the size of the input dimension) that can be effectively treated. The prospect of replicate runs in stochastic simulators introduces a trade-off between replication and exploration, a challenging design issue. The noise, $v$, makes additional demands on the analysis when its variance is non-constant. This article examines these basic issues, identifies accessible and effective methods, and points to unresolved questions that should be addressed.  

Equation \ref{eq:basic} is often used to model physical experiments, where an observation $y(x)$ is truth, $M(x)$, plus measurement error (and, possibly, intrinsic variability as well) or, for an observational study, where $M(x)$ is fit to the observations with residual variance. Because they are structurally the same, physical experiments can be analyzed with methods used for stochastic simulators \citep{gao1996predicting}. However, the contexts and goals are often different, leading to different problem formulations and different interpretations of results.

The choice of method, with its assumptions and limitations, is crucial for any analysis of an experiment. An inclination for simplicity and availability of software would encourage the use of a standard statistical regression model (for example, linear regression) for $M$ with a constant $\sigma_v^2$. That this approach is effective under some circumstances, especially when the space, $X$, of possible inputs is small, begs the question of how reliable it can be as a general prescription. Complex systems modeled by a simulator may neither suggest nor allow much simplification. The methods described in this review allow the simulated data to guide the choice of method under general conditions with little, or no simplification.  Statistics \citep{sacks1989design, kennedy2001bayesian} and Applied Mathematics \citep{sullivan2015introduction} play prominent roles in the design and analysis of deterministic computer experiments. Unsurprisingly, some methods developed for deterministic simulators have modifications that can be used in the stochastic context. Alternatives, driven by the stochasticity, are necessary in many contexts. These structural differences will be noted in the narrative below.

\subsection{Goals}
\label {sec:goals}

We have three primary goals; all related to the cross-disciplinary nature of this topic.

One goal is to bring effective statistical methods to the attention of subject scientists and enable a deeper understanding of stochastic simulators in use. The descriptions below of statistical tools used (or cited) try to avoid being bogged down in mathematical intricacies. Some details of individual methods are included to help in understanding the strengths and weaknesses of the methods. Application of a number of  methods is exemplified on testbed cases (Section \ref{sec:testbed}), and available software for methods are identified where possible.

A second goal is to familiarize statisticians with an area of major importance that is crucial to the formation of evidence-based policy.  Statisticians are sorely needed in the study and application of agent-based models (ABMs) and stochastic simulators in general.  Researchers familiar with deterministic simulation techniques will see immediate opportunities, but statistical expertise of all kinds is essential to advance the study of stochastic simulators.

The analysis of stochastic simulators is a developing field with many unsolved problems.  Challenges are often driven by the scale of the problems and a range of issues whose resolution requires close cooperation between statisticians, subject scientists, and computer scientists. A third goal of this paper is to spur that process.

The review is structured as follows: Section \ref{sec:models} describes the models that form the basis for the analyses; Section \ref{sec:design} is devoted to the fundamental question of what simulator runs to make. Section \ref{sec:calib} addresses a common objective of simulation experiments: calibration. Section \ref{sec:other} discusses other models and objectives that are important, but are more on the ``boundaries’’ of this review and are therefore less detailed. Finally, Section \ref{sec:conclusions} summarizes conclusions and poses unanswered questions. The references here do not cover the entire body of work on stochastic simulators but, together with this overview, should provide adequate coverage of the problems discussed.

\section{Example Simulators}
\label{sec:testbed}

Three stochastic simulators will be discussed throughout this review to aid understanding. Two are deliberately simplified and used to exhibit key features of the methods presented. In some cases simpler strategies could be equally effective because the complexity of the models has been greatly reduced. Since the data/generating mechanisms used are available, others can compare different strategies, but the demonstration purpose is the one that is relevant in the discussion and reported computations.
The third is a model which we use to anchor and motivate methods. The specific model in question is an epidemiological model developed in response to the Ebola epidemic of 2014. For the Ebola model, a synthetic population representing the individuals in Liberia (population $\sim$ 4.5 million) and their activity schedules, inducing a time-varying contact network of individuals and locations, was developed \citep {mortveit2015synthetic}, and paired with an agent-based model \citep{bisset2009epifast}.
Together, this ABM models a contagion spreading from one individual to another in Liberia. Since the parameter for contagion, transmissibility, only controls the \emph{probability} of infection given an interaction occurs, this model (and many like it) is stochastic. The model is updated daily, with the progress of the disease determined by the activity schedule, contact details, and other epidemiological characteristics.  
This model is complex, with high dimensional outputs, multiple unknown inputs, and non-normality all present. The analysis performed by \cite{fadikar2018} tackles all of these using ideas discussed within this article (see Sections \ref{sec:qk}, \ref{sec:multiple}, and \ref{sec:calib} ).

\subsection{Fish Capture-Recapture} 
\label{sec:fish}

The first simplified stochastic simulator we consider mimics the movements and schooling behavior of fish in a mark-recapture application. Mark and recapture involves capturing a sample of the population, marking and releasing them, and following up by capturing another sample and counting how many are marked -- the recaptured. The number recaptured allows estimation of the population size \citep{begon1979investigating}. The process is modeled by initializing a population of fish at random locations in a 2-d, rectangular lake with boundary conditions. The fish begin moving and schooling according to simple, agent-based rules. After an initial period of time, 100 fish are marked as they pass through a ``net’’ in the lake.  After a second period of time, 100 fish are captured using the same net and the number of “recaptured” are recorded.

This agent-based model is a modified version of the flocking model developed in NetLogo \citep{Wilensky1999}. The collective behavior that emerges in the flocking model is the result of providing each individual agent with the same set of simple rules \citep{reynolds1987flocks}. The flocking model is modified to include the mark-recapture dynamics described above. Given an observed count of recaptured fish, this model can be used to estimate the total size of the fish population (see Section \ref{sec:abc}).
The only input considered is the number of fish in the total population and the output is the number of recaptured fish. Other inputs for this model control the individual movement rules of the fish; for simplicity these are ignored here and set to default values. Supplementary code, and compiled Rmarkdown documents, corresponding to our analysis of this simulator can be found at \url{https://github.com/jhuang672/fish}. Running the simulator afresh will require the installation of NetLogo from \url{https://ccl.northwestern.edu/netlogo/}.

\subsection{Ocean Circulation}
\label{sec:ocean}
The second simplified example is a stochastic simulator that models the concentration of oxygen in a thin water layer (around 2000m deep) in the South Atlantic ocean \citep{mckeague2005statistical, Herbei2014estimating}. The physical model is described via an advection-diffusion equation (equation (4) of \cite{mckeague2005statistical}), i.e., a non-linear partial differential equation (PDE) describing the dynamics of oxygen concentration in terms of the water velocities and diffusion coefficients.
For a given set of inputs, the solution of the advection-diffusion equation is not available in closed form. However, using theoretical results \citep{feynman1948, kac1949}, the solution can be closely approximated through an associated random process \citep{Herbei2014estimating}. For a specific location within the domain, random paths of the process are generated, producing noisy outcomes that approximate the solution to the PDE at that location. This example is simplified by taking the oxygen concentration output to only depend on four inputs: two unknown diffusion constants ($K_x$ and $K_y$) and the two location variables (latitude and longitude).  All other inputs are held fixed at nominal values. 
Such stochastic approximations are numerous in physical sciences, either due to computational limitations, a lack of complete understanding of the underlying system, or because the system under study is itself believed to be random.
Supplementary code, and compiled Rmarkdown documents, corresponding to our analysis of this simulator can be found at \url{https://github.com/Demiperimetre/Ocean}.

\section{Statistical Models}
\label{sec:models}

An experiment of running a simulator and producing data whose output is described by equation \ref{eq:basic} can have a multitude of goals. A principal objective, and the one we focus on here, is using the simulated data to predict values of the simulator, $M(x) + v$, and the uncertainties of the predictions, at untried $x$s in a context where getting new runs of the simulator is not cost-free. When $M$ is believed to be “simple” (for example, a polynomial function of the coordinates of $x$) there are many standard “classical” techniques that can be used to approximate $M$. For example, linear regression models and generalised linear models have been used by  \cite{Andrianakis2017efficient} and  \cite{Marrel2012global}. Complex problems such as those in Section \ref{sec:testbed} are less easily managed: specifying a functional form for complex $M$ requires sufficient prior knowledge or a huge abundance of data, both of which are often lacking. A prime emphasis of this article is on methods that have been developed to cope with such concerns; adequate references for a variety of standard methods are available for simpler circumstances.

There are a range of factors that need to be taken into account before choosing a statistical model (hereon referred to as a {\em surrogate model}, as it acts as a surrogate for the computer model). In addition to methodological assumptions, it is important to consider the “context”, that is, the conditions of the particular problem being studied, leading to equation \ref{eq:basic} and its extensions. Some important contexts include:
\begin {itemize}
\item  {The space of inputs is usually a hyper-rectangle: each coordinate of an input $x$ is constrained by upper and lower bounds.   Section \ref{sec:ocean} simplifies issues by taking a rectangular input space even though the Atlantic Ocean is not rectangular}.
\item  {The output $y$ in equation \ref{eq:basic} is scalar, but multiple output, such as time-series, is also common}.
\item  {Some inputs may be categorical rather than numerical}. 
\item  {The probability distribution of $v$, the variability, is often taken to be normal, but often invalid, as with the Ebola model}.
\end{itemize}

Stretching back to \cite {sacks1989design, currin1991bayesian}, a vast literature, mostly on deterministic simulators, has found that a Gaussian Process (GP) model produces a flexible, effective surrogate for $M$. This approach, and its modifications needed to address the presence of input dependent $\sigma_v^2(x)$ in equation \ref{eq:basic}, can be effective for stochastic simulation as has been documented in the literature \citep[e.g.][]{kleijnen2009kriging, kleijnen2017regression} and will be apparent below. A thorough intuitive explanation (for deterministic computer models) can be found in \cite{OHagan2006}. More technical descriptions of GPs from a statistical perspective can be found in \cite{santner2018design} and \citet{gramacy2020surrogates};  for a machine learning perspective, see \cite{rasmussen2006gaussian}. In brief, the use of GPs allows computer model runs to play the key role in selecting a surrogate and assessments of its uncertainty in prediction.  
Deep learning methods, such as neural networks, and other general-purpose predictors are also in wide use. These modern learning machines have difficulties in producing uncertainties and identifying critical inputs but there is active research directed towards that end \citep{neal_bayesian_1996, graves2011practical, welling2011bayesian, papamakarios2019normalizing, gal2016dropout, lakshminarayanan2017simple}.

\subsection{Gaussian Process Surrogates}
\label{sec:gasp}

Suppose that the input space $X$ is a hyper-rectangle in $d$-dimensions; the output $y(x)$ is univariate (scalar); and that variability is normally distributed. Additionally, assume:
 \begin{enumerate}[label=A\arabic*]
\item  The variability, $v$, has constant variance $\sigma_v^2$ 
\item  The mean $M(x) = \mu + Z(x)$
\item  $\mu$ is constant
\item  $Z(\cdot)$ is a Gaussian Process on $X$ with mean 0 and covariance function $K$, deconstructed as a product of a variance $\sigma_Z^2$ and a correlation function $C$.
\end{enumerate}

The technical definition of a GP (Assumption A4) is: for \emph{any} finite $N$ and collection of inputs $X_N = (x_1, \dots, x_N)$, $Z_N = (Z(x_1), \dots, Z(x_N))^\top$ is a multivariate normal random variable  with mean $\textbf{0}$ and $N \times N$ covariance matrix $K_N$, whose entries are $K(x_i,x_j)$. 
It follows that the simulator output, $Y_N = (y(x_1), \dots, y(x_N))^\top$, is also multivariate normal but with mean $\mu\textbf{1}$ and covariance matrix $K_N + \sigma_v^2 I_N$, where $I_N$ is the identity $N \times N$ matrix and $\textbf{1}$ is the $N$-vector of 1s.   

One interpretation is that these assumptions describe a \emph{prior} distribution on all possible functions for the mean $M$. Different choices for the GP allow for different classes of possible $M$; the power of a GP is that these classes can be big enough to allow for all reasonable possibilities. After specifying $\mu$, $K$, and $\sigma_v^2$, a Bayesian analysis can then be carried out, resulting in a \emph{posterior} distribution for all the functions that can still represent $M$ after accounting for the observed simulator runs. 

Another interpretation of $M$ and Assumptions A2 and A4 is to think of $M$ as a random function, with $\mu$ being a regression function (as in linear regression), and the GP for $Z$ modeling the deviation from $\mu$. Both formulations have the same mathematical structure but with differing interpretations.

The predictive distribution for any new run, $y(x_{\mathrm{\mathrm{new}}})$, given the observed simulator data $\{X_N, Y_N\}$ is also normal, and has a known analytical form. The mean $\mu_N(x_{\mathrm{new}})$ and variance $\sigma^2_N(x_{\mathrm{new}})$ of predictions are:
\begin{align}
\mu_N(x_{\mathrm{new}}) &= \mu + k_N(x_{\mathrm{new}})^\top (K_N + \sigma_v^2 I_N)^{-1} (Y_N - \mu \textbf{1}) \label{eq:predict} \\
\sigma^2_N(x_{\mathrm{new}}) &= \sigma_v^2 + \sigma_Z^2 - (k_N(x_{\mathrm{new}})^\top (K_N + \sigma_v^2 I_N)^{-1}k_N(x_{\mathrm{new}}),  \label{eq:varpredict} 
\end{align}
with $k_N(x_{\mathrm{new}})$ denoting the $N$-vector ($K(x_{\mathrm{new}}, x_1), \dots, K(x_{\mathrm{new}}, x_N))^\top$ of covariances between the desired prediction and observed data.
Once the correlation function $C$ is specified, parameters ($\mu$, $\sigma_Z^2$, and $\sigma_v^2$) can be estimated from the data. For specifying $C$, the approach taken for deterministic simulators can be adopted here: specify a parameterized family $C_{\theta}$ and use the data to estimate $\theta$, thereby tailoring $C$ to observations. One example for $C_{\theta}$ is the family of squared-exponential correlation functions (also known as the Gaussian kernel):
\begin{equation}
C_{\theta}(x,w) = {\exp\left\{ - \sum_{j=1}^d \frac{(x_j - w_j)^2} {\theta_j} \right\}}.
\label{eq:corgau}
\end{equation}
This correlation function is suited for approximating very smooth, infinitely differentiable, functions over dimension $d$. Alternative correlation functions exist and are used; one commonly used alternative is the Mat\'ern 5/2 correlation function \citep{stein2012interpolation}, which is appropriate for approximating less-smooth functions (only 2 derivatives).\footnote{The Mat\'ern 5/2 correlation function has the form $\left(1 + \frac{\sqrt{5}(x_j - w_j)}{\rho} + \frac{5(x_j - w_j)^2}{3\rho^2} \right) \exp\left(-\frac{\sqrt{5}(x_j - w_j)}{\rho} \right)$. Further discussion of the features of different kernels can be found in Chapter 4.2 of \cite{rasmussen2006gaussian}, Chapter 2.2 of \cite{santner2018design}, or Chapter 5.3 of \citet{gramacy2020surrogates}.} 
 
With a choice of the family $C_{\theta}$ and Assumptions A1-A4, the likelihood of the observed output is available and maximum likelihood estimates (MLEs) $\hat{\mu}$, $\hat {\sigma_v^2}$, $\hat{\sigma_Z^2}$, and $\hat{\theta}$ can be calculated. Henceforth $\mu_N(x_{\mathrm{new}})$ and $\sigma^2_N(x_{\mathrm{new}})$ will be used to denote the mean and variance of the predictive distribution even when the parameters in equations \ref{eq:predict} and \ref{eq:varpredict} are estimated. The predictive probability distribution for the computer model output $y(x_{\mathrm{new}})$ is then:
\begin{equation}
y(x_{\mathrm{new}}) \sim N(\mu_N(x_{\mathrm{new}}), \sigma^2_N(x_{\mathrm{new}})).
\label{eq:preddist}
\end{equation}
A proper assessment of uncertainty is lost by plugging-in estimated parameters without accounting for their uncertainty. Accordingly, the predictive variance, $\sigma^2_N(x_{\mathrm{new}})$, obtained this way is called the \emph{plugin} (or \emph{nominal}) predictive variance. The alternative of a full Bayesian analysis to estimate the parameters can be computationally impractical in many circumstances, though not impossible (intermediate schemes and approximations have proven to be useful, e.g., \cite{spiller2014automating}).

For the correlation function in equation \ref{eq:corgau}, and for others such as the Mat\'ern 5/2, the correlation between $Z(x)$ and $Z(w)$ depends only on $x-w$, the difference between the two vectors of inputs. That is, $Z$ is assumed to be a stationary GP (and, consequently, so is $y$). For functions exhibiting markedly different behavior in one region of input space than in another part, stationarity is problematic. This issue is tackled and discussed in \cite{gramacy2008bayesian}, \cite{ba2012composite}, \cite{kersaudy2015new}, and \cite{chen2016analysis}, among others, and Section \ref{sec:cattree} discusses one solution.

Despite the fairly complex mathematical expressions above, Gaussian processes are easily accessible thanks to numerous available packages (for example: {\tt DiceKriging} in {\sf R} \citep{DiceKriging}, the {\tt hetGP} {\sf R} package  \citep{hetGP} mentioned later, and the GaussianProcessRegressor function from {\tt scikit-learn} in Python \citep{scikit-learn}). 
In general, a GP is a flexible method for estimating the mean $M(x)$ of the simulator output, despite lack of prior knowledge. This is illustrated in the top panels of Figures \ref{fig:FishHomGP} and \ref{fig:OceanHomGP}, but we first introduce a vital modeling twist to cope with a common feature of stochastic computer simulations.

\subsection{Heteroscedastic GP Surrogates}
\label{sec:hetGP}

The constant variance Assumption (A1) simplifies the construction of a statistical model because only one intrinsic variance parameter $\sigma_v^2$ needs to be estimated. When $\sigma_v^2(x)$ is believed to vary over the input space more must be done. \cite{boukouvalas2014optimal} model $\sigma_v^2(x)$ as $\exp(h(x))$ for simple functions $h$ (e.g., polynomials), a simple extension to assuming just one variance parameter $\sigma_v^2$. (The exponential transform ensures positivity of the variance.) Like analogous approaches to predicting the mean (briefly discussed in Section \ref{sec:gasp}), it isn’t clear what to use for $h$, and its simplicity may not meet the complexities found in many applications.

GPs are used for $\sigma_v^2$ by several authors \citep{Goldberg1998regression, Kersting2007MostLikely, Boukouvalas2009Learning, Ankenman2010, Binois2018JCGS}. The difficulty is that doing so directly depends on observing $\sigma_v^2(x)$ at the inputs $X_n$, but these values are not observed. If there are enough replicated simulation runs, $r_i$, at the inputs $x_i$, then the sample variances ($s^2(x_i) = \frac{1}{r_i - 1} \sum_{j=1}^{r_i}(y(x_{ij}) - \bar {y_i})^2$, for $i = 1 \dots n$) at the $x_i$s can be used to estimate the $\sigma_v^2(x)$ at the inputs $X_n$.  Equations \ref{eq:predict} and \ref{eq:varpredict} can then be used to predict $\sigma_v^2(x_{\mathrm{new}})$. (Working with the logarithm of the sample variances and then exponentiating the results avoids negative predictions of the variance.) But this approach, called {\em stochastic kriging} \citep[SK,][]{Ankenman2010}, is limited by the need for adequate numbers of replicates at each input and the possible inefficiency of treating the variance and mean processes separately.

Those limitations can be removed by considering the intrinsic variances at the inputs, $(\sigma_v^2(x_1), \dots, \sigma_v^2(x_n))$, as unknown parameters (a.k.a., latent variables) to be estimated in the same manner as all the other unknown parameters. \cite{Goldberg1998regression} do so in a fully Bayesian, but computationally taxing, way. Efforts to reduce these costs form the essence of approaches by \cite{Kersting2007MostLikely} and \cite{Boukouvalas2009Learning}. A recent variant, proposed in \cite{Binois2018JCGS} along with accessible software {\tt hetGP} \citep{hetGP}, resolves the computational hazards and is the method described and used in this review.

The technical details addressing the computational barriers of a heteroscedastic GP (hetGP) have three elements. One,  hetGP models the log variances as the \emph{mean} output of a GP on latent (hidden) variables. The second uses Woodbury matrix identities \citep{harville1998matrix} to reduce computations from treating all $N$ observations to computations involving only the $n$ unique inputs, a reduction of computational complexity from $O(N^3)$ to $O(n^3)$, especially relevant when there are many replicates. The third element uses MLE to set all parameters.

While full details are provided by \citet{Binois2018JCGS}, some specifics of the first element of the description above are worth noting.
With $\lambda(x)$ = $\sigma_v^2(x)/\sigma_Z^2$ and $\Lambda_n$  = $(\lambda(x_1), \dots, \lambda(x_n))$ for the $n$ distinct inputs, $\log\Lambda_n$ is taken to be the predictive \emph{mean} of a GP on latent (hidden) variables, $\Delta_n = (\delta_1, \dots, \delta_n)$. For ease of exposition assume the GP has 0-mean (a constant mean is actually the default setting in {\tt hetGP}) and take the covariance function for $\Delta_n$ to be $\sigma_g^2 (C_g + gR^{-1})$ where $g>0$, $R = \textrm{diag}(r_1, \dots, r_n)$, and $C_g$ is a correlation function with parameters $\theta_g$. Then $\log\Lambda_n = C_g(C_g + gR^{-1})^{-1}\Delta_n$. This latent $\Delta_n$ approach facilitates smooth estimates of $\Lambda_n$ and provides a fixed functional form for $\lambda(x)$, but does not incorporate the resulting uncertainty due to the estimates of the intrinsic $\sigma_v^2(x)$ in predictions. Given $\Lambda_n$, the Woodbury identities \citep{harville1998matrix} reduce the likelihood of $Y_N$, the output at \emph{all} inputs including replicates, to depend on quantities of size $n$. Maximum likelihood estimates for the unknown parameters can then be computed at a cost of $O(n^3)$. Derivatives are also computable at a cost of $O(n^3)$, further facilitating optimization for maximizing likelihood.

As a side note, heteroscedastic measurement error is sometimes present in spatial statistics models (which are often related to surrogate models); however we know of no such models which allow for the full modeling and predictions of the intrinsic variance process in the same way as a hetGP. For example, the model in \citet{nguyen2017multivariate} allows for non-constant measurement error at different sites, but it does not estimate these measurement errors jointly with the other model parameters, nor does it allow for the prediction of the measurement errors at new unseen sites. This is mostly because there is little interest in predicting the measurement error process in spatial statistics (the ``true'' underlying signal is the objective), whereas with stochastic simulators the intrinsic variability can be of direct modeling interest.

\paragraph{Fish Example.} We apply both an ordinary homoscedastic GP (homGP) and a hetGP surrogate to the fish example from Section \ref{sec:fish}. The simulation budget is constrained to 400 runs and focuses on the relationship between the total number, $x$, of fish in a population and the number, $y(x)$, of fish recaptured in the second round of capture. The total population is an integer between 150 and 4000. The simulator is run 20 times at each of 20 unique $x$ locations in [150,4000], chosen via a maximin Latin hypercube design (see Section \ref{sec:design}).  The number of fish counted cannot be less than zero, but the normality assumption would allow negative fish counts, so we square root the simulated output before performing our analysis, squaring the resulting predictions to return to the original scale afterwards. In addition we estimate ``truth'' by generating another data set; replicating 500 times at each of the same 20 sites. 

\begin{figure}[hbt!]
\centering
Homoscedastic \hspace{3.6cm} Heteroscedastic \\
\vspace{0.25cm}
\includegraphics[scale=0.4,trim=5 60 25 55,clip=TRUE]{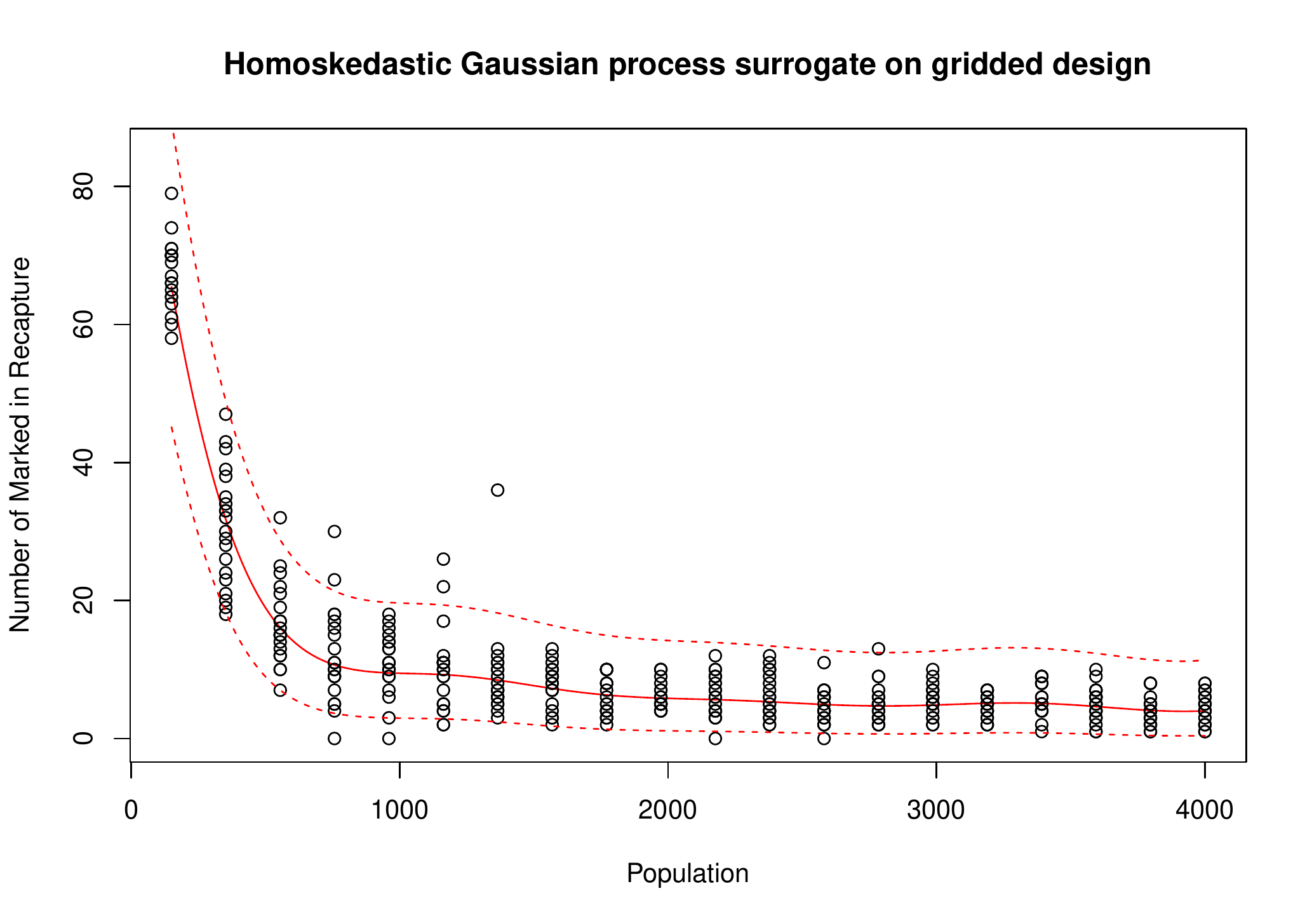}
\includegraphics[scale=0.4,trim=50 60 25 55,clip=TRUE]{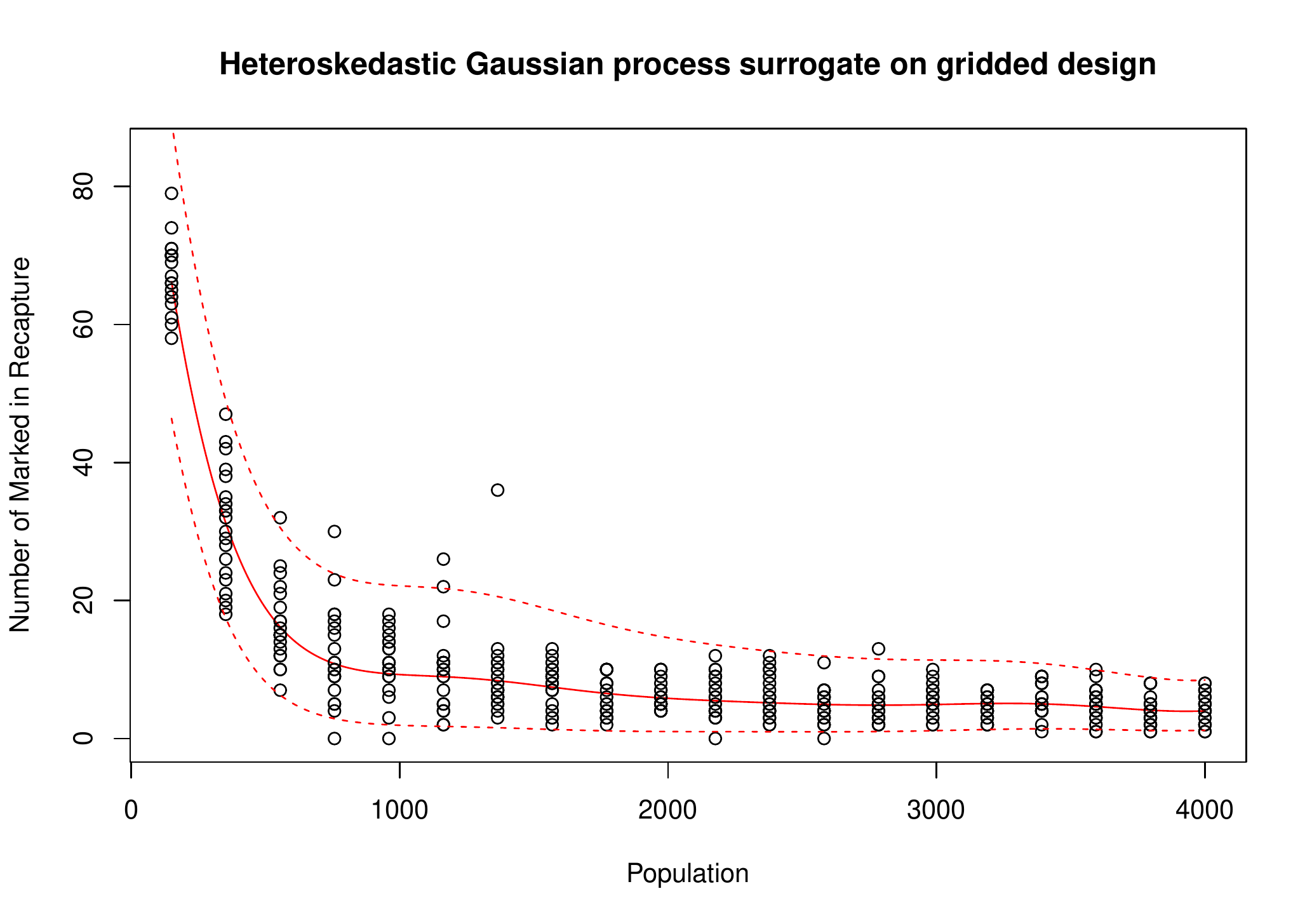} \\
\includegraphics[scale=0.4,trim=5 5 25 55,clip=TRUE]{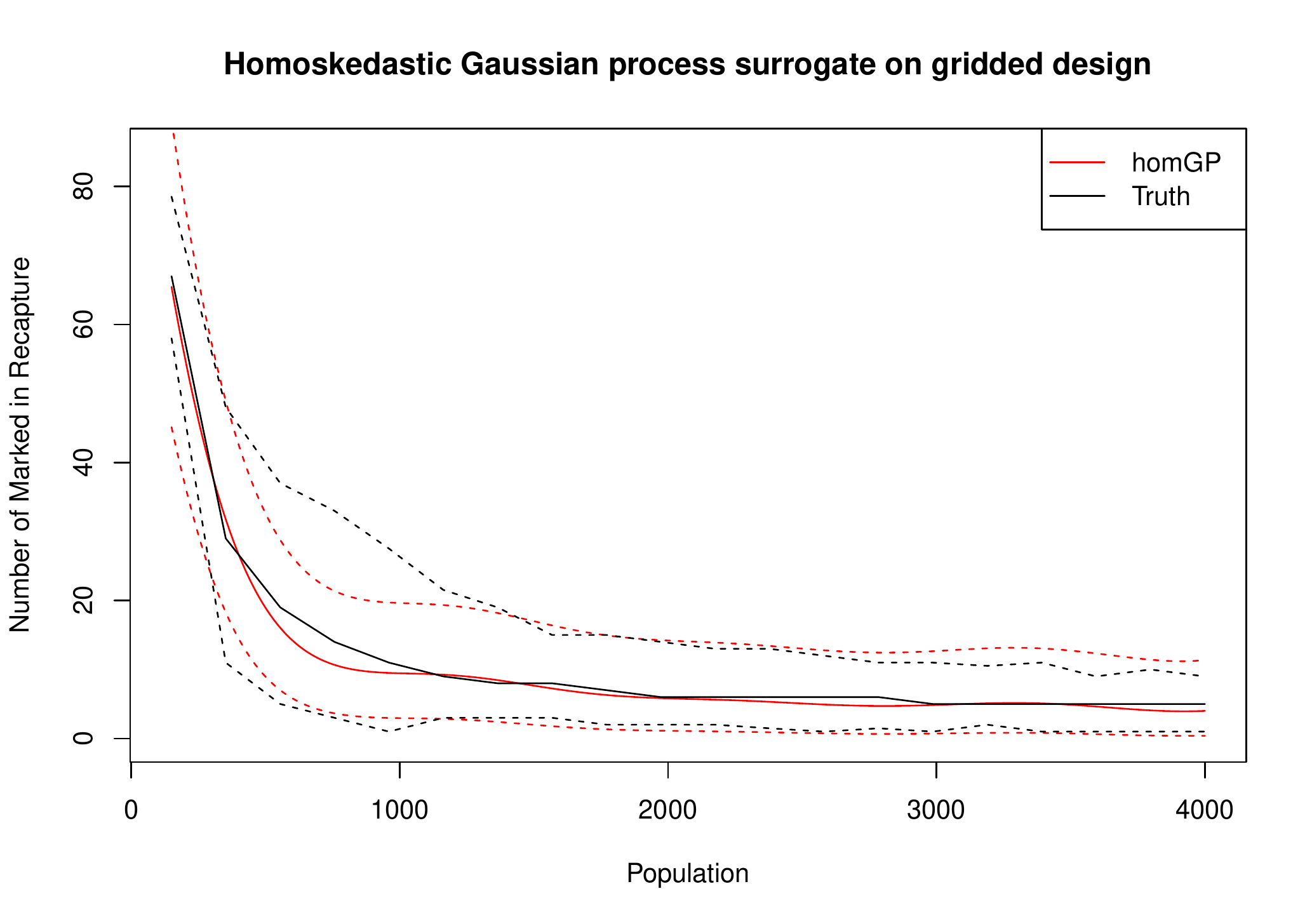}
\includegraphics[scale=0.4,trim=50 5 25 55,clip=TRUE]{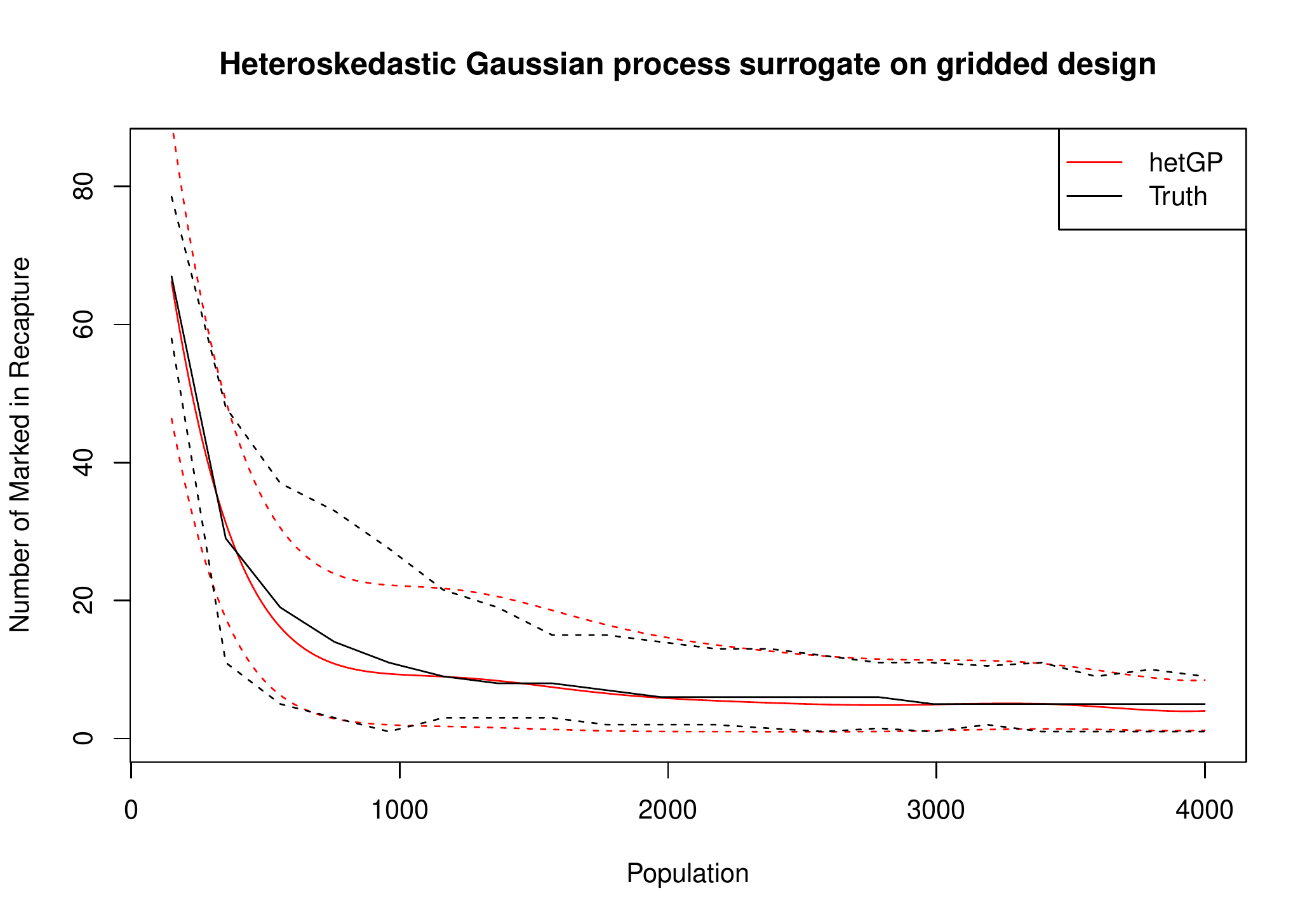}
\caption{Fish example: 400 simulations consisting of 20 replicates at each of 20 population sizes (a maximin latin hypercube scaled to $[150,4000]$, rounded down to nearest integers). The left panels use homGP -- the solid red line is the median of the predictive distribution and the dashed red-lines form the $95\%$ uncertainty intervals. The right panels use hetGP. The upper panels include the data used to fit the surrogates; the lower panels omit the data but include the ``true'' values in black.}
\label{fig:FishHomGP}
\end{figure}

Applying a homGP surrogate with squared exponential correlation function produces the results in the upper left panel of Figure \ref{fig:FishHomGP}; the upper right panel shows the results of hetGP. The predicted intervals for the fish model are obtained in the transformed (square-root) space, and squared to get back to the original space.\footnote{If a large portion of the predictive distribution was negative in the transformed space, the un-transformed intervals would be invalid, but this doesn’t appear to be a problem in our example. Monotonic transforms exist to avoid this problem \citep{johnson2018phenomenological}. Predictions in the transformed space are also provided in the supplementary material.} The lower panels are plots with the ``true'' $2.5\%$, $50\%$, and $97.5\%$ quantiles superimposed.

The key conclusion is that both homGP and hetGP capture the non-linear trend (though a bit off in the region near 800). The presence of non-constant intrinsic variability is clear from the truth plot, with the region near 800 showing higher variability than elsewhere. The hetGP surrogate does not fully resolve the non-constant predictive variability, which includes both the intrinsic variability and that from the surrogate, but does improve on homGP. Full resolution is largely a matter of simulation budget though alternative designs may further improve hetGP.  Our supplementary material includes improved results using the sequential design scheme of Section \ref{sec:seq}. The takeaway message is that the trend is readily treated by both homGP and hetGP; heteroscedasticity encourages use of hetGP perhaps with added simulations or improved designs.

\begin{figure}[hbt!]
\centering
Homoscedastic \hspace{3.6cm} Heteroscedastic \\
\vspace{0.25cm}
\includegraphics[scale=0.45,trim=0 30 60 0,clip=TRUE]{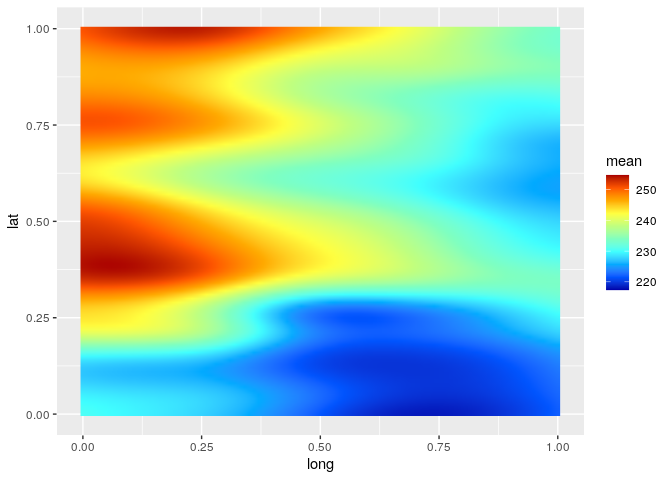}
\includegraphics[scale=0.45,trim=40 30 0 0,clip=TRUE]{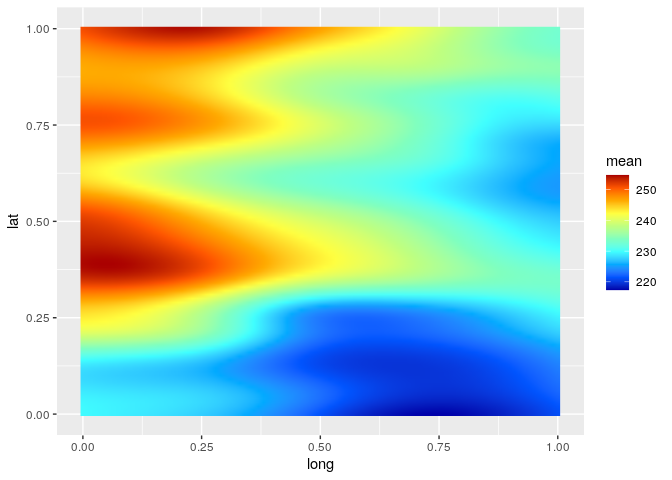} \\
\includegraphics[scale=0.45,trim=0 0 60 0,clip=TRUE]{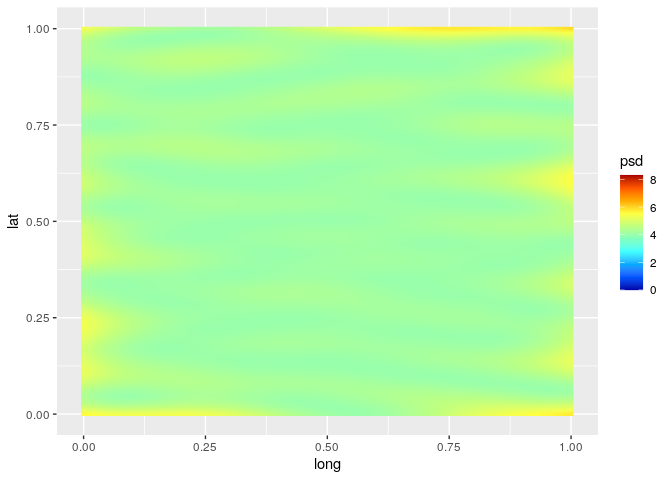}
\includegraphics[scale=0.45,trim=40 0 0 0,clip=TRUE]{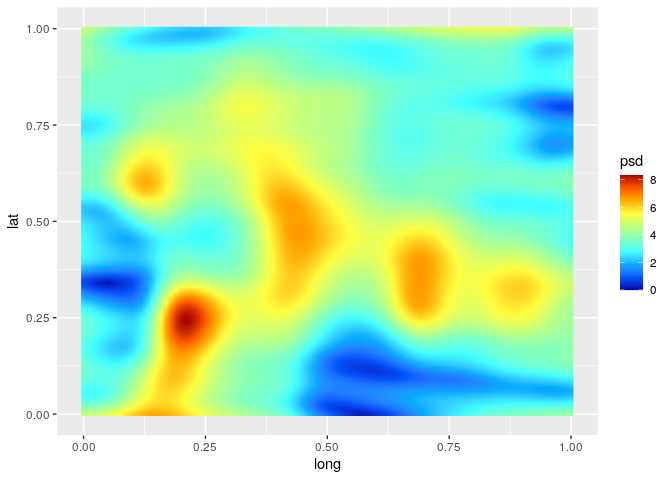}
\caption{Predictive mean and standard deviation surfaces for Ocean output using homGP and hetGP. Data are 1000 simulator runs consisting of 20 replicates at each of 50 input locations from a maximin Latin Hypercube Design (defined in Section \ref{sec:design}) of size 50 in 2 dimensions. The top row provides predictive means, $\mu_N$, and the bottom gives standard deviations, $\sigma_N$, of the predictive distribution of oxygen concentration.  The left column uses homGP, the right uses hetGP.}
\label{fig:OceanHomGP}
\end{figure}

\paragraph{Ocean Example.}  For the ocean model (Section \ref{sec:ocean}), we take each simulation run to be the average of 6 simulation runs. The true simulator is known to be non-normal; this adjustment makes the example more Gaussian. For now, we fix the two diffusion coefficients, $K_x = 700$ and $K_y = 200$, leaving the two spatial coordinates as the only varying inputs. Using 1000 simulations (50 sites each replicated 20 times), we obtain, for surrogates homGP and hetGP, the predictive mean surface and the predictive standard deviation surface (that is, the standard deviations for prediction of the simulator output, accounting for both the uncertainty around the predictive mean and the intrinsic variance estimate $\sigma_v^2$). These surfaces are plotted in Figure \ref{fig:OceanHomGP}, with the left column for homGP and the right column for hetGP.

The mean surfaces for both surrogates are similar. The predictive standard deviation for homGP (bottom-left) is relatively constant across the input area (clearly affected by the constraint that the intrinsic variance $\sigma_v^2$ is constant). The standard deviation surface for hetGP is markedly different, evidence that intrinsic variance is non-constant. The ``truth'' is obtained using replicate runs (up to 100,000) of the simulator at 500 sites (chosen via a LHD, Section \ref{sec:design}), averaging the replicates at each site to get the true mean and the square deviations from the mean to get the true variance). These are plotted in the appendix (Section \ref{sec:appendix_truth}) and the supplementary material; they confirm the presence of non-constant intrinsic variance. Moreover, the standard deviation plot for hetGP exhibits a structure similar to the truth plot, leading to the conclusion that hetGP is the better surrogate for this problem. However, this conclusion comes with a caution: repeating this experiment reveals a great deal of variability in the standard deviation plot, due to variability in the design and the simulations (discussed further in Section \ref{sec:seq}).

Overall, reliable predictions of the mean are achieved, but the uncertainties are less certain. This is similar to the the Fish example, and improving the uncertainties would require more simulation. These results point to the superiority of hetGP to homGP. This is confirmed via a numerical comparison in Section \ref{sec:seq}, where a sequential design is also examined and compared.

\subsection{Non-Normal Variability}
\label{sec:nonnormal}

In many applications, assuming the variability $v$ to be normally distributed is inappropriate. For example, count data, as in the Ebola or fish model, is non-normal and cannot be less than 0. Additionally, the distribution of $v$ at a given input $x$ may not be unimodal: in the Ebola example, even with the inputs $x$ fixed, repeated simulations can lead to two distinct groups of possible infection counts, implying bimodality. In some simulators, there may be a greater tendency for extreme values (fatter tails) in the distribution of $v$. With these possibilities normality can be a strong assumption to be used with caution.

Transformation of the data is a time-honored device that sometimes induces ``enough'' normality in the data to 
permit the use of Gaussian-based methodology (as in Section \ref{sec:hetGP} for the fish model). For example, \cite{henderson2009bayesian} uses the logit transformation ($\log {y/(1-y)}$) in analyzing the proportion of deletions
in mitochondrial DNA.  \citet{Plumlee2014building} take a different route by focusing on the quantiles of the output distribution — normality is not needed. Both of these 
approaches have the appeal of leading to relatively simple modifications of the methods in Sections \ref{sec:gasp} and \ref{sec:hetGP}. 

There are also more complex methods that generally lack the 
same ease of implementation. For example, \cite{Moutoussamy2015} attempt to model the underlying probability density function itself, rather than the output $y$. \cite{Xie2017} devise a Student $t$-process that is not much different than the GP process while at the same time allowing heavier tails in the distribution of the data.\footnote{The \texttt{hetGP} package also implements a Student-$t$ variant \citep{wang2017extended,shah2014student,chung2019parameter}.}

\subsubsection{Quantile Kriging}
\label{sec:qk}

Quantile Kriging (QK) is an increasingly popular tool for the emulation of stochastic computer models \citep{rannou2002kriging,Plumlee2014building,zhang2017asymmetric,fadikar2018}. These approaches are a natural extension of spatial kriging formulations \citep{zhang2008loss,zhou2012estimating,opitz2018inla} used in environmental applications, often with modeling further tailored to account for rare events and extreme quantiles.

The QK method directly models specific quantiles of interest, such as the median and the lower/upper $95\%$ quantiles at each input. Minimal assumptions about the distribution of the simulator output are required. $Q_q(x)$, the $q^{th}$ quantile of the simulator output at input $x$, is modeled with a GP. Given values $Q_q(x_i)$ at inputs $x_1, \dots, x_n$, the quantile, $Q_q{\mathrm (x_{new}})$ for $\mathrm x_{new}$, can be predicted using equations \ref{eq:predict} and \ref{eq:varpredict}. This framework allows the distribution of the variability $v$ to take on almost any shape. Although a true generative process for the output $y$ is lost, we can describe its distribution. 

To implement QK, values of the targeted quantiles at the inputs are needed. Just as in Section \ref{sec:hetGP}, where sample variance estimates at the inputs can be used, sample quantiles can be used here. Said sample quantiles are calculable given enough replicates $r_i$ at each $x_i$. The GPs used to predict new quantile values, $Q_q(x_{\mathrm{new}})$, should also include a noise term $\sigma_q^2$ to acknowledge that the sample quantiles are estimates. Assuming the variability of the sample quantiles is normally distributed may also be invalid, but is at a level further removed from the quantity of interest, $y$, and is often acceptable in practice.

Including the quantile $q$ as an additional input to the GP model can be a useful modification. The quantile $Q_q(x)$ can be reformulated as $Q(x, q)$, increasing the dimensionality of the inputs from $d$ to $d+1$. This strategy allows for the prediction of $Q(x, q)$ for any desired quantile $q$, not just those that were empirically estimated, and is used by \cite{fadikar2018} for the Ebola model.

Alternative QK-based approaches are also under development. For example; a promising variant of QK called Asymmetric Kriging \citep[AK,][]{zhang2017asymmetric} does not require sample quantiles by leveraging quantile regression methods \citep{koenker1978regression}.

\paragraph{Fish Example.} For the fish simulator, QK is implemented with the same simulated dataset as before since many replicates are available. The sample $5\%$, $27.5\%$, $50\%$, $72.5\%$ and $95\%$ quantiles at each of the 20 population sizes form the observed data, and the modification using the quantile $q$ as an added input dimension is adopted. Figure \ref{fig:FishHomGP} presents the predicted $Q(x,q)$ mean for 5 different quantiles along with the data (the left plot) and compares the ``true'' values with predictions at the $5\%$,  $50\%$, and $95\%$ quantiles (the right plot).

\begin{figure}[hbt!]
\centering
\includegraphics[scale=0.4,trim=5 5 25 55,clip=TRUE]{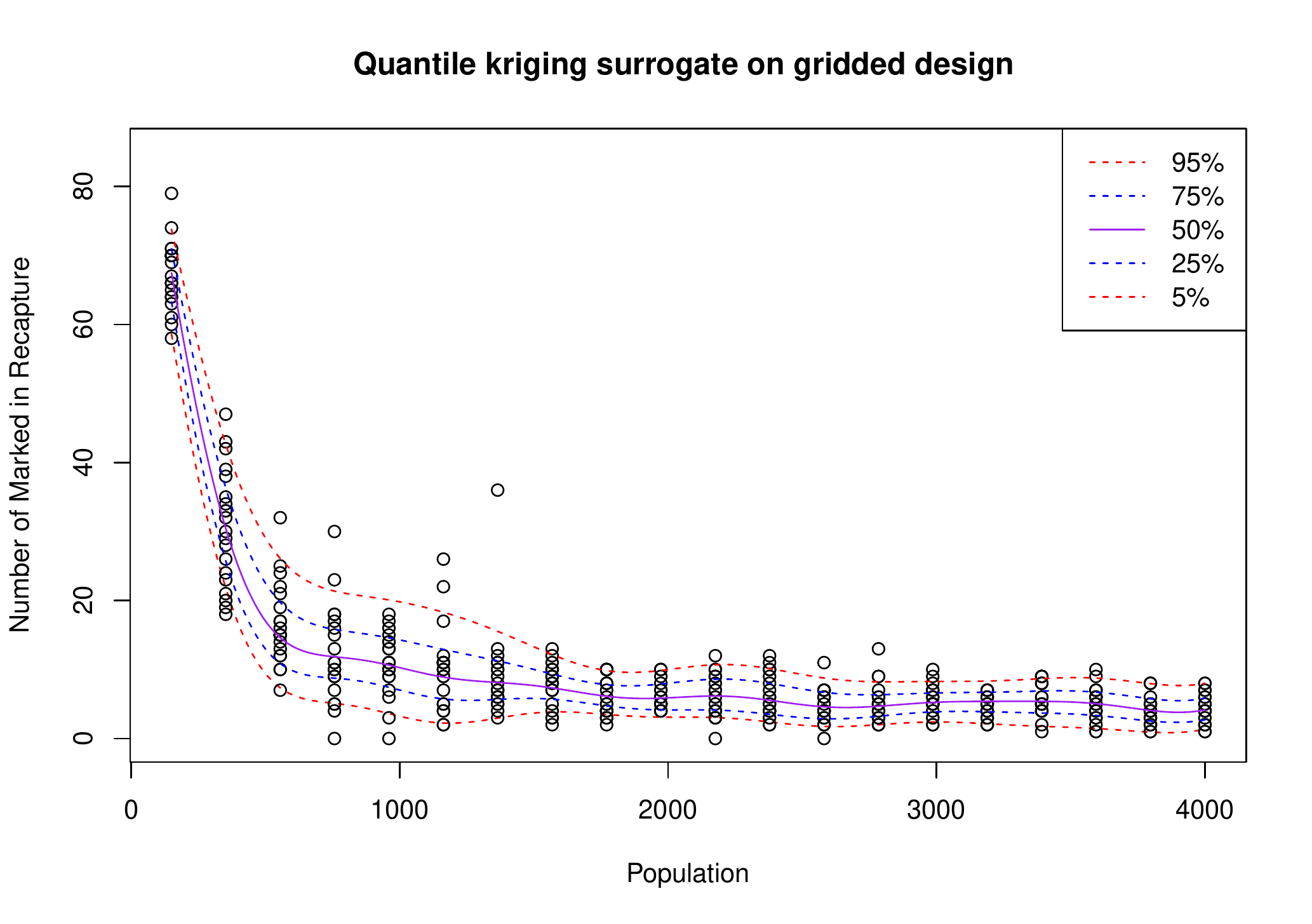}
\includegraphics[scale=0.4,trim=50 5 25 55,clip=TRUE]{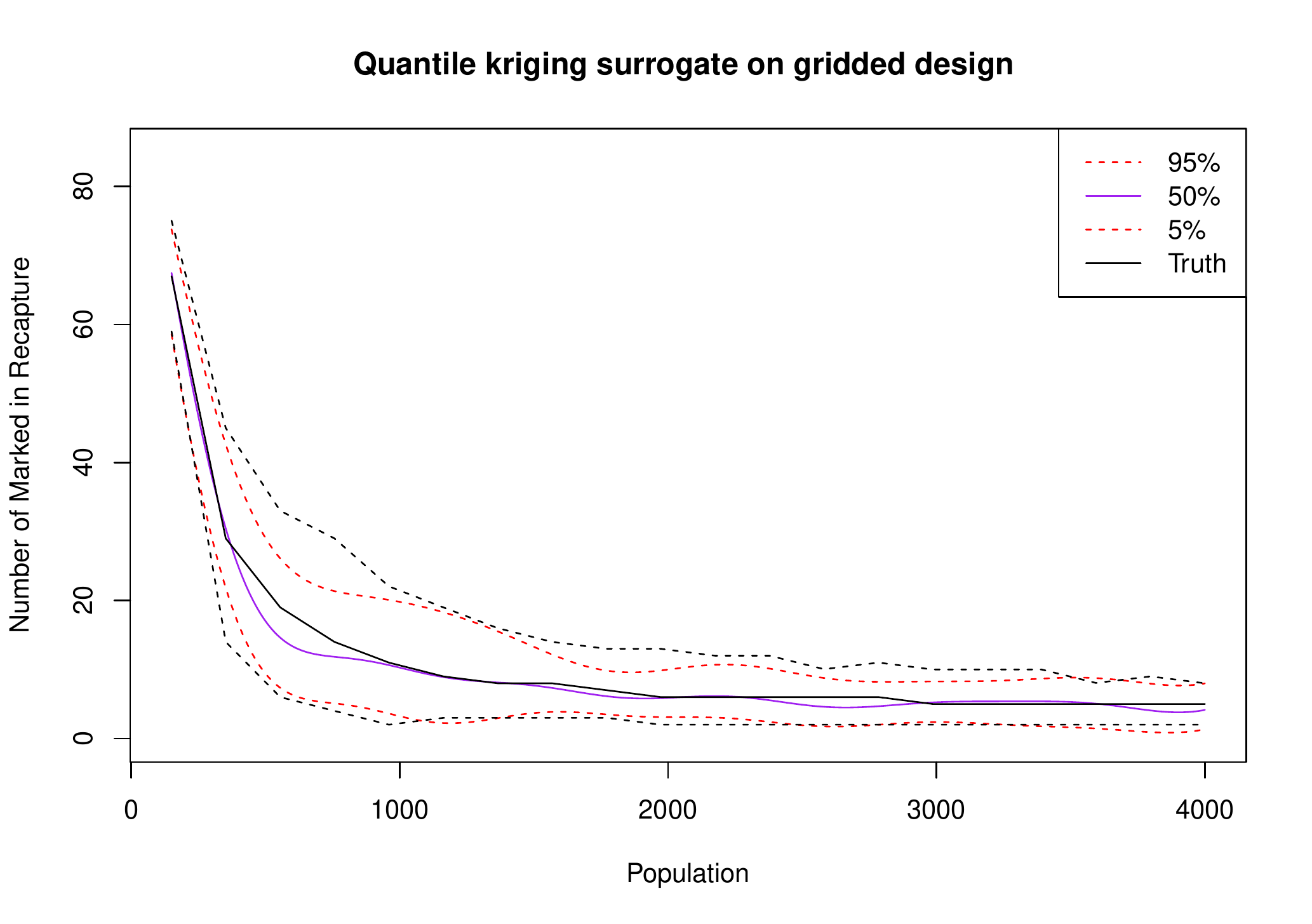}
\caption{Same setup as Figure \ref{fig:FishHomGP} but with a QK surrogate. Mean predictions of 5 quantiles ($5\%$, $25\%$, $50\%$, $75\%$ and $95\%$) are provided on the left along with data, and mean predictions of 3 quantiles ($5\%$, $50\%$, and $95\%$) are provided on the right along with the ``true'' values.}
\label{fig:FishQK}
\end{figure}

The center purple curve in Figure \ref{fig:FishQK} is the predicted median. The outer red lines are the predicted $5\%$ and $95\%$ quantiles; the inner blue curves are the predicted $25\%$ and $75\%$ quantiles. The non-monotone ``wavy’' lines for the $5\%$ and $95\%$ quantiles reflect the natural variability of extreme quantiles based on only 20 observations. Without an abundance of replicates, accurately capturing extreme quantiles is difficult, a drawback of QK. The other quantiles presented display more regularity.

The results from QK do not differ much from those in Figure \ref{fig:FishHomGP} where the square-root transformation was sufficient. With more complex problems, such as the Ebola model, the method is suitable while other approaches may be less so. In any case, QK can be a good robust choice given adequate data for estimating quantiles.

\subsection{ Multiple Outputs }
\label{sec:multiple}

The discussion thus far has assumed that the simulator outputs a single scalar quantity of interest. For multivariate output a more comprehensive model would be ideally used. Sophisticated approaches building multivariate GPs have been employed \citep{conti2010, fricker2013, paulo2012calibration}. Bespoke, problem-specific formulations for time-series and other outputs have also been entertained \citep{Farah2014bayesian,sun2019synthesizing}.
If there are a small number of outputs, treating each independently, with its own surrogate model, often suffices. This method can be effective, despite ignoring any correlation between the different outputs and thus wasting information. For example, \citet{spiller2014automating} deploy independent surrogates at each of a multitude of sites in a region to good effect.

Alternatively, by treating the index, $t$, of the $T$ outputs as an additional \emph{input} dimension (changing the dimension of the input space from $d$ to $d+1$) a GP surrogate on $d+1$ dimensions can be formed \citep{bayarri2009predicting}. This method allows correlation structures between the different outputs to be modeled. This is similar to the QK modification where quantile levels are treated as an added input (Section \ref{sec:qk}). A drawback of this technique is that, if $T$ is very large, computational issues will arise because the GP must be trained on $NT$ data points rather than just $N$.  Intrinsic variability prevents simplifications of the sort used in \cite{bernardo1992integrated} for deterministic simulators in this setting.

A different approach reduces the effect of the size of $T$ to a smaller $K_0$ by representing the output through the use of basis functions, $\psi(t)$:
\begin{equation}
y(x,t) = \sum_{k=1}^{K_0} w_k(x) \psi_k(t) + \delta(x,t).
\label{eq:pc}
\end{equation}
Coefficients $w_k(x_i)\  k=1, \dots, K_0$ are determined by the data; and $\delta(x,t)$ is the residual error between the basis function representation and the data $y$. If $K_0 = T$ then $\delta = 0$. Typically, $K_0$ is taken to be much less than $T$ but large enough so that the error, $\delta$, is sufficiently small. Each $w_k(x)$ can be independently modeled with a surrogate and predictions for $y(x,t)$ are obtained from equation \ref{eq:pc}, ignoring $\delta$.

Different choices for the bases can be appropriate in different settings. For example, \cite{bayarri2007computer} use wavelets for the $\psi_k$s in a deterministic setting where $t$ is time. A common choice of basis functions are principal components: the $\psi$s are the eigenvectors of the matrix $Y_N^\top Y_N$, the first $K_0 $ of which are in correspondence with the first $K_0$ eigenvalues in decreasing order. It is often the case that the first few (five or less) principal components are enough to capture sufficient information about the full ($T$) data set. Coefficients $w_k(x_i)$ are then equal to  $\sum_{t=1}^{T} y(x_i,t)\psi_k(t)$. More information about principal components can be found in \cite{jolliffe2011principal} and software for obtaining  $\psi$ and  $w_k$ is prevalent.

Further discussion about using principal components to model high-dimensional simulator output can be found in \cite{higdon2008computer}. Principal components are also utilized in \cite{fadikar2018} to model the time-series output of the stochastic Ebola simulator. While principal components are a common default, there is concern that key features of the data set may be left within the discarded $\delta(x,t)$ preventing reliable prediction. \cite{salter2019uncertainty} document these concerns with regards to calibration and suggest an alternative.

For problems with functional outputs, with potentially missing data and/or irregularly spaced data (such as irregularly spaced timesteps or spatial locations), a functional decomposition can also be useful. For example, \cite{ma2019computer} use functional principal component analysis to model satellite observation simulations.

\section{Experimental Design}
\label{sec:design}

For an experiment, the design (the choice of $x$ values) and analysis (the assessment of the output $y(x)$) are, in principle, closely connected. Other considerations can also enter. For physical experiments, controlling for external influences or nuisance factors by blocking and randomization is often a vital part of the design. External influences are absent in computer experiments and so controlling for nuisance factors is usually irrelevant. However, many minor parameters are often fixed which could instead be randomized over, with a consequent addition to intrinsic error. 

With a specific goal (e.g., predicting simulator output) and a criterion of accuracy  (e.g., the average prediction uncertainty: the integrated mean-squared prediction error, IMSPE\footnote{With $\sigma^2_N(x)$, the predictive variance, the IMSPE, of a design $D$ is equal to $\int_{x \in X} \sigma^2_N(x)\, dx$.}), designs that optimize the criterion are preferred. Since the criterion will usually depend on the surrogate, which, in turn, depends on unknown parameters, what to use as a stand-in for the parameters before any data are collected is an issue. Extensive study of single-stage deterministic computer experiments resolved this dilemma by downplaying optimality and recommending readily computed ``space-filling’' designs where no large region of input space is missed. Space-filling designs are readily computed, whereas optimizing IMSPE is complicated and without substantial advantage. For practical adoption, designs must be easy to produce as well as effective.

Multiple methods exist for obtaining space-filling designs, the most popular being Latin hypercube designs \citep[LHDs;][]{mckay1979comparison}.\footnote{A Latin hypercube design is one where: on each dimension, the input space is divided into, usually, equal intervals and each interval is constrained to contain exactly one data point.} LHDs have proved adequate, especially when joined with an additional criterion, such as the maximin criterion, where one also maximizes the minimum distance between points in the design.\footnote{Such maximin LHDs are purportedly produced for example, by the maximinSLHD  function of the {\sf R} package {\tt SLHD} \citep{SLHD}, or the lhs function from the Python package {\tt pyDOE} \citep{pyDOE}.} Even a random LHD will often suffice.
Sobol sequence designs \citep{sobol1967distribution} are equally effective for predicting the output of a deterministic simulator. The $x$s for Sobol designs are generated sequentially making it easy to retain the space-filling character when a multi-stage or sequential design strategy is used.\footnote{In {\tt R}, the sobol function in the {\sf R} package {\tt randtoolbox} \citep{randtoolbox} can be used to generate Sobol sequences.}
\cite{Pronzato2011} have a lengthy discussion of these and other space-filling methods, some pertinent to non-rectangular geometries.

For stochastic simulators the picture is far less clear. The presence of intrinsic variability raises the complication of replication, not present in deterministic experiments. With the same inputs, a stochastic simulator can be run multiple times (replicated) providing different output values each time due to the intrinsic randomness. Replicates obviously have an effect on the estimation of the intrinsic variance, $\sigma_v^2$, and therefore on prediction (see Section \ref{sec:hetGP}), and so the number and location of replicates are important. A simple approach for a single-stage experiment is to use a space-filling design to establish the sites $X_n = (x_1, \dots, x_{n})$ of the experiment and then add replicates at each site. Determining the number, $r_i$, of replicates at each site $x_i$ and how to apportion between replicates and sites, that is, how to choose the number of unique sites, $n$, given a total simulation budget $N$, is not well understood. In fact, there is limited theoretical evidence of the need for replicates altogether, although there is numerical evidence and wide belief that replicates can be advantageous, at least in appropriate contexts. For example, \cite{wang2019controlling} produce designs by minimizing bounds on IMSPE.  Their numerical results show no need for replicates unless $\sigma_v^2(x)$ is large compared to $\sigma_Z^2$ (a factor in measuring uncertainty in estimating the mean $M$). 

The presence of intrinsic variability suggests there is value in multi-stage designs where stage 1 is used to get information about  $\sigma_v^2(x)$ and later stages exploit this information to allocate replicates and select new inputs. Questions arise as to how inputs should be selected for stage 1, and also how to leverage the results from stage 1 to select new inputs and replicates in later stages. The two factors, replication and multiple stages (including fully sequential), are central to developing adequate design strategies. Attention is paid to both factors in the discussion below.

\subsection {Single-Stage Design}
\label{sec:onestage}

A common approach in single-stage studies is to use space-filling designs for inputs, say $n$ in number, and $r$ replicates at each input, sometimes with no repeats i.e., $r=1$.  Predictions follow as described in Section \ref{sec:models} depending on the particular prediction model selected. Choices have to be made about the total number of runs and the number of replicates at each input site ($N=nr$). Often, $N$ is a question of budget, but there is little insight into how $r$ should be chosen except when meeting a specific surrogate model requirement, as in SK (Section \ref{sec:hetGP}). 

For their single-stage study, \cite{Marrel2012global} use a standard LHD with no repeats to compare the performance of different statistical models. On the other hand, \cite{Plumlee2014building} use a LHD with varying numbers of replicates $r_i$ at each $x_i$. In their case, the number of replicates \emph{must} be large, because the QK method (Section \ref{sec:qk}) depends on computing  quantiles of the output $y(x_i)$ at each input site of the design.

\subsection {Two-Stage Design}
\label{sec:twostage}

The case for a two-stage design is largely to enable estimation of $\sigma_v^2$ at stage 1 and use it for the second stage. \cite{Ankenman2010} provide one solution in the context of SK. A first-stage design chooses the $x_i$s via an LHD of size $n_1$ with a common number, $r$, of replicates at each of the inputs, resulting in a total number of $N_1 = n_1 r$ runs at stage 1. The first-stage analysis uses the $r$ replicates at each input to estimate $\sigma_v^2(x_i)$ using the sample variances. As outlined in Section \ref{sec:hetGP}, a GP (working with $\log s^2(x_i)$) is then used to produce a “plug-in” estimate of $\sigma_v^2(x)$ for all $x$. A different GP uses that variance estimate to build a predictor for the mean output $M$.

For stage 2, $n_2$ additional unique input locations are chosen so that the combined set of design locations, $X_n =(x_1, \dots, x_n)$, remains space-filling. The IMSPE is then calculated by integrating the MSPE all possible inputs $X$, using the GP model constructed in stage 1. Minimizing the IMSPE with respect to the number of replicates $R_n = (r_1, \dots, r_n)$ provides the optimal number of replicates for the chosen $X_n$. Details are in \cite{Ankenman2010}. One difficulty is that the optimal $R_n$ might produce an $r_i$ for a first-stage site that is smaller than the $r$ already used at stage 1. Some fix to the method would then be necessary.

In this setting, a Sobol sequence could be used to obtain a design that is space filling at both stage 1 and stage 2. This is not what is done in \cite{Ankenman2010}, but a Sobol sequence is easier to implement and likely to yield similar results. Choosing the unique inputs $X_n$ for stage 2 by optimizing the IMSPE could also be done, but adds to the computational burden. Suitable recommendations for the values of $n_1, n_2, N$ and the replicates at each distinct input are lacking (in \cite{Ankenman2010} the recommendations are ad hoc) and, as for one-stage experiments, open for study.  A third-stage design (or indeed, any multi-stage design) can be constructed by repeating stage 2 in the above process.

\subsection{Sequential Design}
\label{sec:seq}

When the statistical design and resulting analysis are closely coordinated, it may be feasible to carry out a sequential process whereby, after the first stage, a run is chosen one-at-a-time. After each run all quantities of relevance can be updated in order to determine the next run. This addresses the issue of learning about $\sigma_v^2$ and obtaining new runs without pre-specifying their allocations. An advantage of a sequential design is the possibility of stopping when a criterion is met before a budget constraint is reached.  Another advantage is the increased likelihood of making useful runs of the simulator, replicates or otherwise. For some objectives, such as optimization (Section \ref{sec:opt}), a sequential design is usually essential. For global prediction, \cite{Binois2018TECH} present an approach to sequential design, implemented in the previously mentioned {\tt hetGP} package. 

The strategy in \cite{Binois2018TECH} begins at stage 1 with a space-filling design $D_1$ of $n_1$ inputs and an allocation of runs $(r(x_1), \dots, r(x_{n_1}))$. Using a GP for $M$ and a latent GP prior on $\sigma_v^2$, as in Section \ref{sec:hetGP}, a MLE computation deals with all parameters, leads to predictors, and a calculable estimate of IMSPE($D_1$). A new point $z$ is considered, either as a new unique input $x_{{n_1}+1}$ or as a replicate of an existing input in $D_1$. Selection $z$ is added to the design $D_1$ if $z$ minimises IMSPE($D_1 + z$), yielding a new design $D_2$. This myopic rule can be iterated and each time a new point is added the surrogate, including MLEs of its parameters is updated. The process stops when a criterion is met or the computational budget exhausted.

Computational viability is strained by the updating required after each run. On the other hand, the computational burden is eased by nature of it being ``greedy’': it only seeks the optimal data point for the very next simulator run, ignoring runs that may be better in the long run. 

This is not the only sequential design scheme available for global prediction problems. For example, the tree-generating processes used in TGP and BART (see Section \ref{sec:cattree}) deliver specialized sequential design strategies. Details are available in \cite{gra:lee:2009} and \cite{chipman2010bart}.  

Blurring the lines between multi-stage and sequential designs, it can sometimes be practical to run additional simulations in batches (e.g., as in making efficient use of a multi-core supercomputer). In such circumstances a ``batch design'' would be desirable. These have been developed for deterministic experiments \citep{Loeppky:2009,Duan:2017,Erickson:2018}, but not yet explicitly extended to stochastic cases.

When fully sequential methods are feasible the seqhetGP strategy sketched above is valuable. There are several aspects worth examining: 

\begin{itemize}
\item The extensive use of a surrogate in the construction of the design requires scrutiny by diagnostics that assess the quality of the surrogate.    

\item  The first stage of a sequential strategy must avoid a poor (e.g., too small) initial design lest a poor starting surrogate leads to poor choices thereafter.

\item The utility of a sequential design depends on the relative cost of implementation compared to simulator runs. For challenging problems simulator runs are likely to be costly enough to make sequential design attractive.

\item There may be modifications to a sequential design that reduce computational load without paying a significant cost in accuracy. For example, re-estimate parameters periodically rather than after each step.

\end{itemize}

\begin{figure}[hbt!]
\centering
\includegraphics[scale=0.45,trim=5 0 5 0]{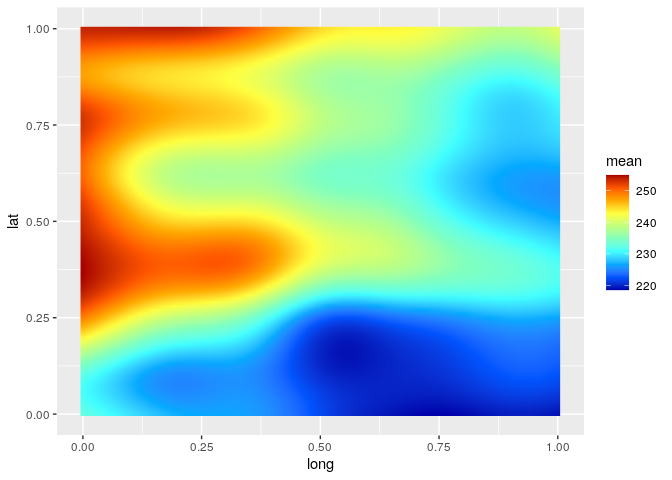}
\includegraphics[scale=0.45,trim=40 0 5 0,clip=TRUE]{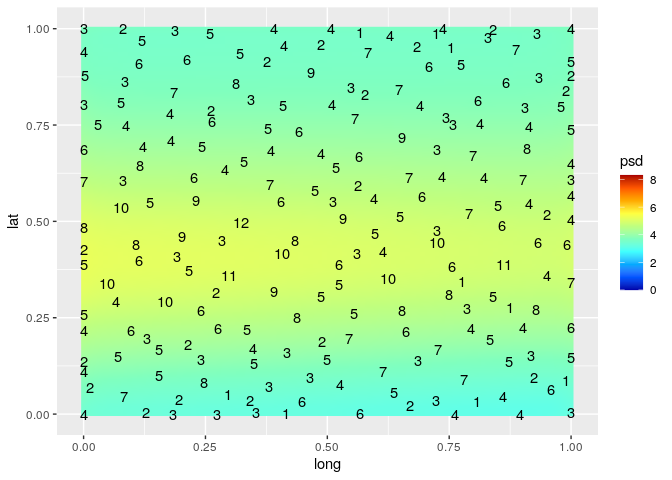}
\caption{Ocean Prediction with seqhetGP. Left plot is the mean, $\mu_N$, of the predictive distribution; right plot the standard deviation, $\sigma_N$. Design sites with their replicates are superimposed on the right-hand plot.}
\label{fig:OceanSeqGP}
\end{figure}

\paragraph{Ocean Example.} For the ocean model, we use an initial design of 50 sites, chosen by a maximin LHD in 2-$d$, each site with 5 replicates. The remaining 750 data points are then assigned via the sequential scheme. The resulting mean and standard deviation surfaces are in Figure \ref{fig:OceanSeqGP}. For the standard deviation surface the design sites are superimposed along with the number of replicates taken at the sites.
The mean surface in the left panel is slightly different than for the non-sequential analyses (Figure \ref{fig:OceanHomGP}, top row). The standard deviation surfaces look very different. For the design itself, new inputs are heavily replicated in regions where the standard deviation is large, and less so in regions where it is small. Additionally, the sequential design includes more unique sites than the fixed design, and more points on the boundaries of the input space.

Using the “truth” established in Section \ref{sec:hetGP} we can compare the performances of the three methods. As discussed previously, the visual presence of heteroscedasticity is a deterrent to using homGP. Visually distinguishing between the performances of the hetGP and seqhetGP surrogates is more difficult: the means appear similar, and whilst some patterns in the true standard deviation appear to be captured by hetGP, imperfections are visible and the magnitude is not always correct. With the seqhetGP standard deviation, nuance seems lost. To properly compare the different methods, a numerical comparison can be more valuable.

Two useful numerical measures are root mean squared error, RMSE (the square-root of the average squared difference between the surrogate's prediction of the mean and the ``true'' mean) and Score (the proper scoring rule from equation 27 in \cite{gneiting2007strictly}). RMSE measures the accuracy of the mean predictions and Score is an overall measure testing the accuracy of the combined mean and variance predictions. With a test set of inputs ${x_1, \dots, x_p}$ and simulator outputs ${y_1, \dots, y_p}$, surrogate predictive means ${\mu_N(x_1), \dots, \mu_N(x_p)}$ and variances ${\sigma^2_N(x_1), \dots, \sigma^2_N(x_p)}$, Score is
\begin{equation}
    \frac{1}{p}\sum_{i = 1}^{p} \left(- \left(\frac{y_i - \mu_N(x_i)}{\sqrt{\sigma^2_N(x_i)}}\right)^2 - \log(\sigma^2_N(x_i))\right).
\end{equation} 
Smaller RMSE is better while for Score, larger is better.

For the three methods, the RMSE for homGP, hetGP and seqhetGP are respectively 2.056, 1.985, and 1.567; and the Scores are respectively -3.999, -3.880, and -3.834. The RMSE results reveals that seqhetGP is best at predicting the mean, which was not obvious from the plots. The Scores for hetGP and seqhetGP are close but noticeably better than homGP, affirming the presence of  heteroscedasticity. 

\begin{figure}[hbt!]
\centering
\includegraphics[scale=0.8,trim=5 0 5 0]{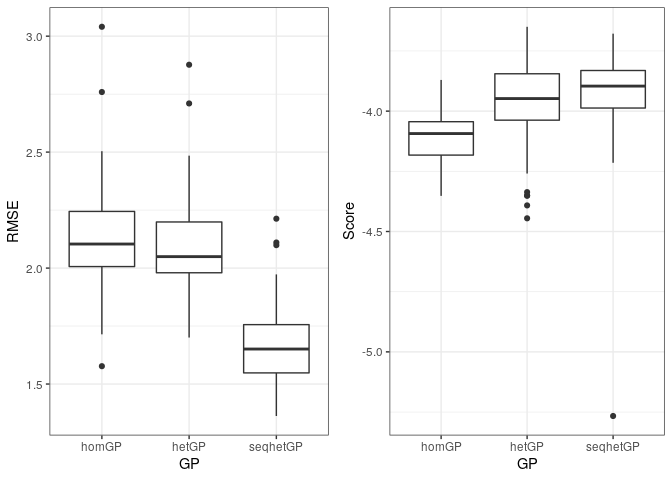}
\caption{Performance results of the three ocean model surrogate model fits, repeated 100 times: the boxplots are for RMSE and Score computed, for each repetition, at the 500 test locations.}
\label{fig:boxplots}
\end{figure}

The randomness in stochastic simulators as well as variability in design (there are many possible maximin LHDs) can induce a large degree of variability in specific results such as those just cited. It is therefore difficult to rely on a single result for making comparisons. As such, the above experiment is repeated 100 times and the resulting 100 RMSEs and Scores are summarized in boxplots in Figure \ref{fig:boxplots}.

The boxplots confirm what was found with the single data set: heteroscedasticity is present and seqhetGP is preferred. Visual inspection of many of the standard deviation plots for the repeated experiments (as discussed in Section \ref{sec:hetGP} and found in the supplement) reveals considerable variation and departure from the true standard deviation. The variance (the intrinsic variance and the GP uncertainty for the mean) can be hard to get right without an abundance of data, and the difficulty is compounded by the use of plug-in estimates whose uncertainty is not accounted for. 

\subsection{Designing for Statistical Model Parameter Estimation}
\label{sec:otherdes}

Sections \ref{sec:twostage} and \ref{sec:seq} construct designs that rely on a surrogate model based on stage 1 data in order to choose subsequent data points. The quality of the designs depends on the accuracy of the surrogate  which, in turn, depends on the accuracy of its parameters. An alternate approach to those used in Sections \ref{sec:twostage} and \ref{sec:seq}, is to construct an initial design with the express purpose of better estimating these parameters. 

\cite{boukouvalas2014optimal} address the problem and focus on hetGP models, using a simple parametric function for the variance ($\sigma_v^2(x) = \exp(h(x))$, where $h$ is a simple function (e.g., a polynomial). They propose designs that maximize a criterion previously used for deterministic simulators by \cite{abt1998fisher}: the logarithm of the determinant of the Fisher information matrix, $\log{|I|}$. Numerical results suggest this method gives improvements in estimating the parameters, but overall global prediction is no better, and sometimes worse, than using a space-filling design. When prediction is of prime importance, the question arises about how to make use of such designs for stage 1 in a multi-stage or sequential setting, where its impact on obtaining better initial surrogate models can be felt. For example, see \cite{zhang2018distance}.

\section{Calibration}
\label{sec:calib}

Calibration is needed when there are inputs to the simulator that are neither known nor measurable, a common condition in practice. Transmissibility in the Ebola simulator and the diffusion coefficients in the ocean model are examples of such inputs. In order to infer (indirectly) values for these inputs and produce predictions, added information in the form of field data (experimental or otherwise) are necessary. Inclusion of field data and calibration parameters, labelled $u_C$, leads to the observation model:

\begin{equation}
y_F(x)  = y_S(x,u_C) + \delta_{\mathrm{MD}}(x)  + \epsilon,
\label{eq:obsmodel}
\end{equation} 
where $y_F(x)$ are real-world field observations at controllable (or measurable) inputs $x$, $y_S$ is the simulator with additional unknown, non-measurable, inputs $u_C$, $\epsilon$ is measurement error for the observations $y_F(x)$ (with variance $\sigma_{\epsilon}^2$), and $\delta_{\mathrm{MD}}(x)$ is an important term that accounts for the simulator not being a perfect representation of reality. $y_F$ ``observes'' reality with error $\epsilon$;  reality = $y_S+\delta_{\mathrm{MD}}$. 

Multiple competing methodologies and even philosophies exist for calibration. Several solutions to the calibration problem are outlined below.  Despite the centrality of calibration in computer experiments, comprehensive comparisons are lacking.

\subsection{Kennedy-O’Hagan Calibration (KOH)}
\label{sec:KOcalib}

The formulation in equation \ref{eq:obsmodel} was made by \cite{kennedy2001bayesian} for deterministic simulators and is the basis for much of the calibration and related prediction work since. The strategy pursued by \cite{kennedy2001bayesian}, as implemented in \cite{bayarri2007}, obtains a surrogate for $y_S$ and models $\delta_{\mathrm{MD}}(x)$ with a GP (although other choices are possible). After replacing $y_S$ with the surrogate, posterior distributions for all unknowns can be obtained via a Bayesian analysis. In practice, the surrogate model is fit only using the simulator data, ignoring possible influences from the field data. Details and discussion of this modular approach can be found in \cite{bayarri2007} and \cite{liu2009modularization}.

The KOH approach emphasizes the necessity to address calibration and model discrepancy together. Confounding between $u_C$ and $\delta_{\mathrm{MD}}(x)$ inevitably occurs because there are multiple combinations of $u_C$ and $\delta_{\mathrm{MD}}(x)$ that result in the same observed field data. Thus, $u_C$ is non-identifiable and its estimation is compromised, as is the discrepancy. Nonetheless, the resulting \emph{predictions} for $y$ and $E(y)$ are sound, even if the individual estimates for $u_C$ and $\delta_{\mathrm{MD}}(x)$ aren't. For details and further discussion see \citet{higdon2004combining}, \citet{bayarri2007}, \citet{bryn2014learning}, and  \citet{tuo2016theoretical}. 

Multiple attempts to circumvent confounding have surfaced. \cite{tuo2015efficient} alleviates the ambiguity in $u_C$ by formally defining it as a least-squares quantity; \cite{gu2018scaled} propose novel priors for the discrepancy that compromise between the \cite{tuo2015efficient} strategy and KOH; and \cite{plumlee2017bayesian} introduces priors on the discrepancy that are orthogonal to the prior mean. In the stochastic simulator literature, \cite{Oakley2017calibration} removes $\delta_{\mathrm{MD}}$ but compensates by inflating the variability in the prior distribution for $u_C$. Ignoring $\delta_{\mathrm{MD}}$ altogether can be justified by strong evidence of the simulator being accurate, but such evidence is rare.

For stochastic problems, where reality is stochastic the discrepancy term $\delta_{\mathrm{MD}}(x)$ cannot be assumed deterministic. Modeling the discrepancy is likely be influenced by the model for the simulator while recognizing that discrepancy is often smoother. For example, if modeling $y_S$ calls for a hetGP with the Matern 5/2 correlation function then it is likely that a hetGP is needed for the discrepancy, perhaps with the smoother squared exponential correlation. A full Bayesian analysis in such circumstances may be prohibitively expensive and the above procedure would have to be modified. \cite{sung2019calibration} use a hetGP for the discrepancy (but for a deterministic simulator), estimating parameters via maximum likelihood and following \cite{tuo2015efficient} to avoid confounding.

\paragraph{Revisiting Ebola.} The Ebola study \citep{fadikar2018} calibrates an ABM using the KOH framework. The simulator $y_S$ has 5 unknown, unmeasured inputs $u_C$ and the output is the log of the cumulative number of infected individuals up to week 1 and every week thereafter up to 57 weeks. The field data $y_F$ is a set of reported cumulative counts. For the statistical model, a QK strategy (Section \ref{sec:qk}) is followed by replicating each distinct simulation 100 times and then condensed into evenly-spaced quantiles at each time point (specifically, the $5\%$ $27.5\%$, $50\%$, $72.5\%$ and $95\%$ quantiles). These quantile output trajectories are then reduced to a more manageable 5 dimensions using the principal component decomposition outlined in Section \ref{sec:multiple}. 

Underlying the approach is an assumption that the epidemic trajectories (actual and simulated) can be approximated by quantile trajectories (i.e., a realized epidemic that resembles the $q^{th}$ quantile at time 1 will also resemble the $q^{th}$ quantile at a later time).
Accordingly, the quantile $q$ is included as an input parameter (see Section \ref{sec:qk}) to allow KOH calibration to learn about the 5 calibration parameters as well as the (unknown) value of $q$ for the observed epidemic. Because the difference between the simulator and reality quantile trajectories could not be noisy, the discrepancy is treated as deterministic (a smoothing spline is used rather than a GP for the discrepancy). Posterior distributions for unknown $u_C$, $\delta_{\mathrm{MD}}$, and $q$ are obtained and used to make predictions of the cumulative counts and other quantities.

In the main analysis, which restricts the field data to only the first 20 weeks, the estimate of model discrepancy is almost zero. A subsequent analysis done using field data up to week 42 exposes some inaccuracy of the simulator (non-zero $\delta_{\mathrm{MD}}(x)$) — the simulator continues to predict infections, even after the epidemic has died in reality.

\paragraph{Ocean Example.} 

The previous ocean analyses fixed the two diffusion coefficients. Realistically, they are unknown and calibration is necessary. ``Field'' data are artificially created by averaging over 200 simulations at 150 different longitude-latitude coordinates, using the previously fixed values of the diffusion coefficients ($K_x=700$ and $K_y = 200$). ``True'' values are obtained by adding a fake discrepancy, taken as a single realization from a GP with a squared-exponential correlation function, a variance of $1.64$, and $\theta$ values of (1, 2) (equation \ref{eq:corgau}). To these, normally distributed pretend ``observation errors'' with a variance of 4 are added,  two such observations at each site. In real problems, the field data would be observed and not generated like this. Note that field data for this problem corresponds with the mean of the simulator, not individual draws from the simulator; a result of the simulator being a stochastic approximation.

With the diffusion coefficients now uncertain, the simulator has four inputs. A computer experiment is designed with runs at the 150 sites used for the field data and 500 unique selections of the calibration parameters $K_x$ and $K_y$. This is done by combining copies of the 150 longitude and latitude sites with a size-500 maximin LHD for ($K_x$, $K_y$), and then improving the combined design by maximizing the minimum distance between design points in the 4-dimensional space. Call this set of points $D_{\mathrm{oc}}$. The simulator experiment is carried out by taking 10 replicates at each point in $D_{\mathrm{oc}}$. Two distinct surrogates (a homGP and a hetGP) are fit with this fixed design. In addition, a seqhetGP surrogate is constructed, with an initial design of only 4 replicates of $D_{\mathrm{oc}}$ and the remainder of the budget assigned following the strategy of \cite{Binois2018TECH}, described in Section \ref{sec:seq}.      

\begin{figure}[hbt!]
\centering
\includegraphics[width=\linewidth]{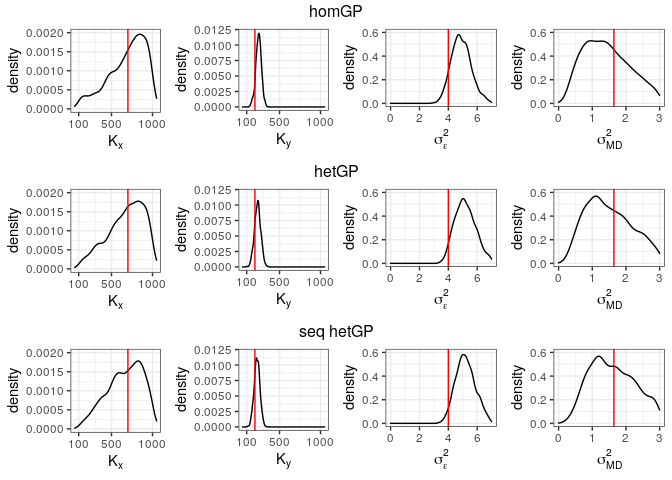}
\caption{Calibration results for the ocean model. The top row shows the posterior densities for the four parameters using homGP with the fixed design, the middle row uses hetGP with the fixed design, and the bottom row uses the hetGP surrogate with the sequential design. The budget for all three is 5000 runs. True values are superimposed as red vertical lines.}
\label{fig:OceanCalibration}
\end{figure}

For a KOH analysis done in modular fashion the surrogates are fit only using the simulated data. Because reality here is represented by the expectation of the simulator (rather than the simulator output itself), $y_S$ in equation \ref{eq:obsmodel} is replaced with $E(y_S)$. Similarly, because reality is deterministic, $\delta_{\mathrm{MD}}$ is modeled as a standard GP. Of course, the simulated data are outputs from $y_S$, not from $E(y_S)$ — the surrogate is used to approximate the deterministic $E(y_S)$. MCMC is then used to obtain posterior distributions for the remaining unknowns: the diffusion coefficients, $K_x$ and $K_y$; the variance and correlation parameters of the model discrepancy GP, $\sigma_{\mathrm{MD}}^2$ and $\theta_\mathrm{{MD}}$; and the observational error, $\sigma_{\epsilon}^2$). The posterior distributions for the key parameters are in Figure \ref{fig:OceanCalibration}; their true values are $K_x = 700, K_y = 200, \sigma_{\mathrm{MD}}^2 = 1.64$ and $\sigma_{\epsilon}^2 = 4$.

For all three surrogate models the posterior distributions for $K_x$ are fairly diffuse. The $K_y$ posteriors are highly concentrated, but not quite around the true value. The three posteriors for observational error are quite similar but all point to estimates closer to 5 rather than the true 4. The posteriors for the discrepancy variance are diffuse. These plots underline the dilemma of calibration: obtaining accurate values of calibration (and other) parameters in the presence of model discrepancy is problematic. Additionally, with noisy data, it is difficult to obtain precise estimates. However, KOH does yield useful posterior predictive distributions. 

Table \ref{tab:OceanCompareCalib} compares predictions by KOH calibration with 3 other calibration approaches. The first one estimates $K_x$ and $K_y$ by ordinary least squares (OLS): $(K_x, K_y)$ is chosen such that the sum of the squared residual difference between the mean surrogate prediction and the observed data is minimized. New observations are then predicted by running the surrogate with the parameters $(K_x, K_y)$ replaced by the OLS estimates $(\hat{K}_x, \hat{K}_y)$. The second approach follows a frequently adopted practice by guessing, or ``judiciously selecting'', specific values for $K_x$ and $K_y$. Here, the choices $K_x = 600$ and $K_y = 400$ are made, and then predictions are made using the surrogate.  Call this method SINGLE. The third method, NOCAL, generates predictions as if there were no field data and the distribution for $(K_x, K_y)$ is taken as their prior distribution, independent uniform priors on [100, 1000]. In these alternative methods, the observational error variance is fixed at the true value, and 0-discrepancy is assumed (the former is overly generous and the latter is all too common in practice). For NOCAL, a distribution for the oxygen concentration is obtained by sampling values of $K_x$ and $K_y$ from their prior distribution and plugging them into the surrogate, while for KOH, by sampling from the posterior distributions of all unknowns.

\begin{table} [!ht]
\centering
\begin{tabular}{| c | c | c  c  c  c  c| }
\hline
 & & $(\hat{K}_x, \hat{K}_y)$ & OLS & SINGLE & NOCAL & KOH \\
 \hline
 \multirow{3}{*}{RMSE} & homGP & (824.9, 295.4) & 9.16 & 9.16 & 9.22 & 9.14\\
 & hetGP & (754.9, 295.8) & 9.16 & 9.16 & 9.20 & 9.15\\
 & seqhetGP & (496.3 276.0) & 9.15 & 9.15 & 9.22 & 9.16\\
 \hline
 \multirow{3}{*}{Score} & homGP &  & -2.50 & -2.66 & -2.59 & -2.32\\
 & hetGP &  & -2.55 & -2.71 & -2.62 & -2.32\\
 & seqhetGP &  & -2.55 & -2.69 & -2.61 & -2.30\\
\hline
\end{tabular}
\caption{Performance results of the three ocean model surrogates under KOH calibration. RMSE at the 500 test locations with the ``true'' values used for Figure \ref{fig:boxplots}; similarly for Score. $(\hat{K}_x, \hat{K}_y)$ are least squares estimates for $(K_x, K_y)$, OLS presents the predictive results from least squares calibration, SINGLE the results from arbitrarily choosing $(600, 400)$ for the diffusion coefficients, NOCAL the results from sampling the prior for $(K_x, K_y)$, and KOH the results from performing KOH calibration.}
\label{tab:OceanCompareCalib}
\end{table}

Although the differences in RMSE are negligible, the Scores indicate that KOH performs the best. It is also possible that the accuracy of OLS, SINGLE, and NOCAL is overstated, because the observational error variance is taken as known while in KOH it is estimated. That the least squares estimates $(\hat{K}_x, \hat{K}_y)$ are not always close to the true values is unsurprising given the presence of discrepancy, along with possible imperfections and high variability in the surrogate. For similar reasons, scant differences appear among the three surrogates. 

The similarity of RMSEs is a consequence of large variability in the surrogate, the presence of discrepancy, the dominance of the longitude and latitude inputs, and a weak effect from the calibration inputs. The first explains the magnitude of the RMSEs and the last explains why fairly inaccurate calibration inputs (in OLS and SINGLE) don’t matter. Because KOH addresses discrepancy, its Score exceeds the others’, showing that accounting for the discrepancy is necessary and can’t be wished away

\paragraph{Estimating Calibration Parameters}

The focus here, consistent with Section \ref{sec:design}, has been on improving global prediction. If the problem was instead to provide good estimates for calibration parameters when model bias is absent, then different designs may be better suited. \cite{damblin2018adaptive} address this in the context of deterministic simulators, but it is unclear how such methods extend to stochastic simulators. Additionally, whilst KOH facilitates capable predictions, the complexity and recorded pitfalls in KOH has led to competing calibration techniques that are also in common usage.

\subsection{History Matching (HM)}
\label{sec:HM}
	
History Matching (HM) is a common alternative to KOH calibration \citep{craig1997pressure,vernon2010galaxy, Boukouvalas2014Calibration, Andrianakis2017efficient}. HM searches for inputs where the simulator outputs closely match observed data, while recognizing the presence of the various uncertainties, including model discrepancy. The HM approach rules-out ``implausible'' inputs in a straightforward way, rather than attempting to find probable inputs. With an observation $y_F$, and initially assuming $u_C$ makes up all the unspecified simulator inputs, $u_C$ is deemed implausible if:
\begin{equation}
\frac{|y_F - \mu_N(u_C)|}{\sqrt{\sigma^2_N(u_C) + \sigma^2_{\mathrm{MD}} + \sigma^2_{\epsilon})}} \geq 3,
\label{eq:eqhm}
\end{equation}
where $\sigma^2_N, \sigma^2_{\mathrm{MD}}$, and $\sigma^2_{\epsilon}$ are the variances of the surrogate, the model discrepancy, and the observational error respectively. In other words, an input is implausible if the difference between the observation and the simulator output using that input is sufficiently large relative to those uncertainties. The number 3 comes from \cite{pukelsheim1994three} who shows that at least $95\%$ of \emph{any} unimodal distribution is contained within three standard deviations. When there are multiple outputs or additional, controllable inputs there are modifications to equation \ref{eq:eqhm} \citep{vernon2010galaxy}. 

The process can be repeated in so-called ``waves'', using non-implausible $u_C$ found at one wave to generate simulation runs for the next wave, sequentially reducing the space where $u_C$ could lie. With these waves HM aims to avoid regions of inputs where $u_C$ is unlikely to be and, in that regard, HM is a calibration design scheme. At any given wave, it is possible for all values of $u_C$ to be deemed implausible — the so-called terminal case  \citep{salter2019uncertainty} — usually implying that $\sigma^2_{\mathrm{MD}}$ is set too low or that the simulator is not fit for purpose.
\cite{Andrianakis2015bayesian} contains a thorough description of HM whilst applying it to a complex epidemiology model of HIV.

\paragraph{HM and KOH.}  With KOH the estimation of $u_C$ is confounded with discrepancy, but predictions and their uncertainties are available. However, implementing KOH in complex problems may be burdensome if not intractable. Speculatively, a hybrid strategy may be to use HM to reduce the input space, confirm the absence of the terminal case, and then apply KOH in the narrowed space to get predictions and uncertainties. Complex models, unlike the fish and ocean examples in this review, would be ones for which this approach would be most appealing. Such hybrid strategies are a topic for further exploration.

\subsection{Approximate Bayesian Computation (ABC)}
\label{sec:abc}

Obtaining posterior distributions for calibration parameters $u_C$ and predictions (e.g., the KOH approach in Section \ref{sec:KOcalib}) can be computationally challenging. ABC methods offer an alternative which have been found useful in moderately complex contexts \citep{rutter2018microsimulation}; but less so in more ambitious settings \citep{Mckinley2018approximate}. 

ABC aims to produce samples from $\pi(\theta | Y_F)$, the posterior distribution of  unknowns $\theta$, given the field data $Y_F$. For calibration, think of $\theta$ as $u_C$. ABC does this by generating samples for the unknowns $\theta^{(s)}$ and the output $z^{(s)}$ from $\pi(Y_F | \theta)\pi(\theta)$, that is, from the likelihood of the data given the unknowns, multiplied by the prior probability of the unknowns. For computer models, generating samples from the likelihood is equivalent to running the simulator. Such samples are only accepted if $z^{(s)} = Y_F$. For continuous settings, where exact equality cannot occur, acceptance is instead made if $B(z^{(s)}, Y_F) < \tau$, where $B$ is a measure of distance and $\tau$ a level of tolerance. An approximated posterior distribution is then given by the collection of accepted $\theta^{(s)}$s. When there are multiple outputs (or there are other controllable inputs $x$, and so for any given $\theta^{(s)}$ there are effectively multiple outputs), $Y_F$ and $z^{(s)}$ can be replaced with informative summary statistics.  Finding a single statistic sufficient for all outputs is challenging, and a poorly chosen one can invalidate results. 

The choice of the tolerance $\tau$ is important.  If $\tau$ is small then it may take a very long time to generate a single sample that satisfies the inequality. If $\tau$ is not small then the approximation to the posterior is less reliable. For calibration, $\tau$ can be interpreted as a bound on the observational error and model discrepancy, leading to a ``correct'' posterior rather than an approximation \citep{wilkinson2013approximate}. This is then similar to HM with the subjective choice of bounds. 

ABC can be done without the use of a surrogate, but many runs of the simulator itself to generate many $\theta^{(s)}$ may be required. Otherwise, too few accepted $\theta$ will remain, or an overly high value of $\tau$ will be required. In either case accuracy can be compromised. Such computational barriers can be alleviated by the use of a surrogate.

\paragraph{Fish Example.} 

\begin{figure}[hbt!]
\centering
\includegraphics[scale=0.35,trim=25 5 0 10]{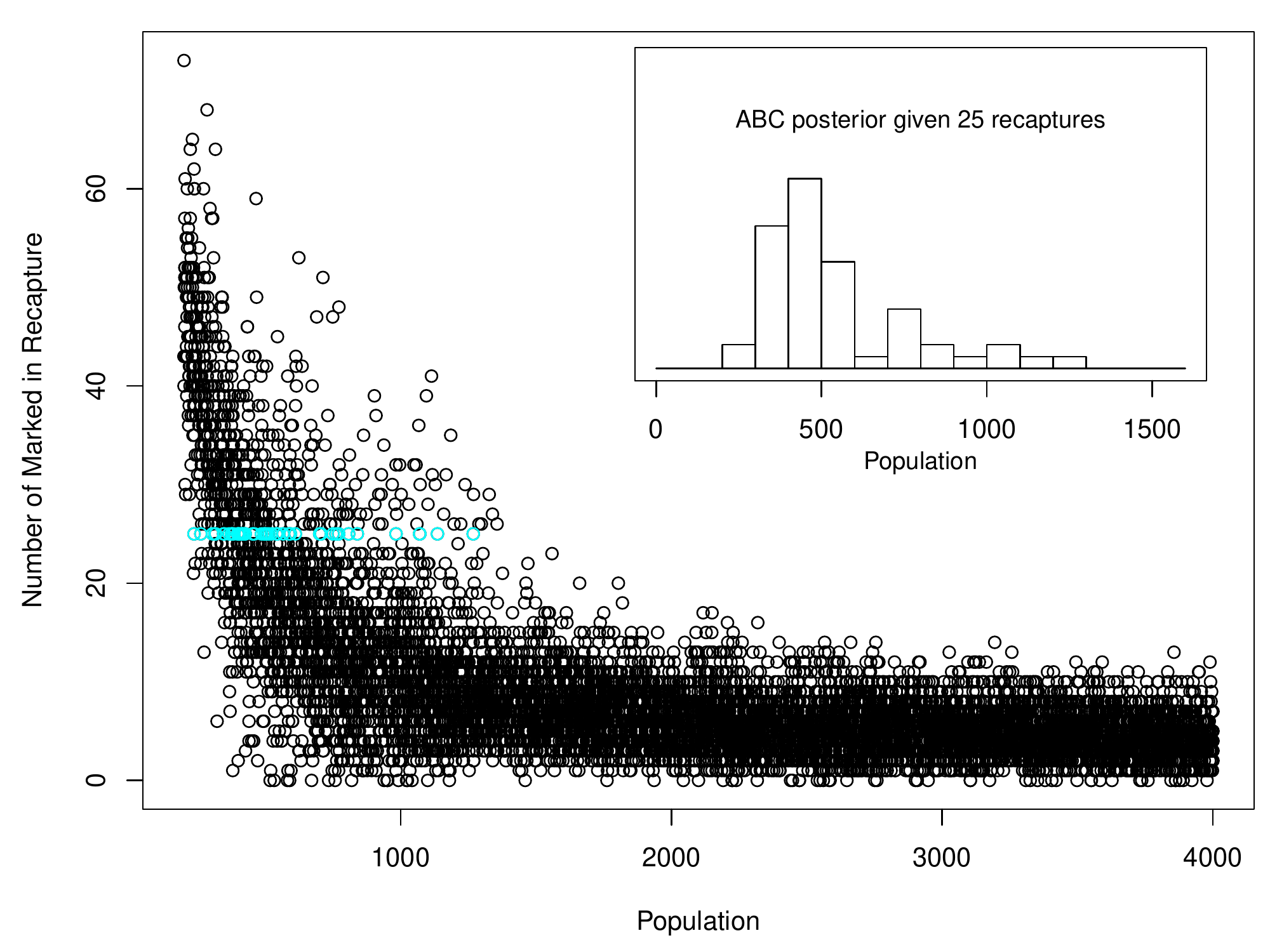}
\includegraphics[scale=0.35,trim=10 15 20 10]{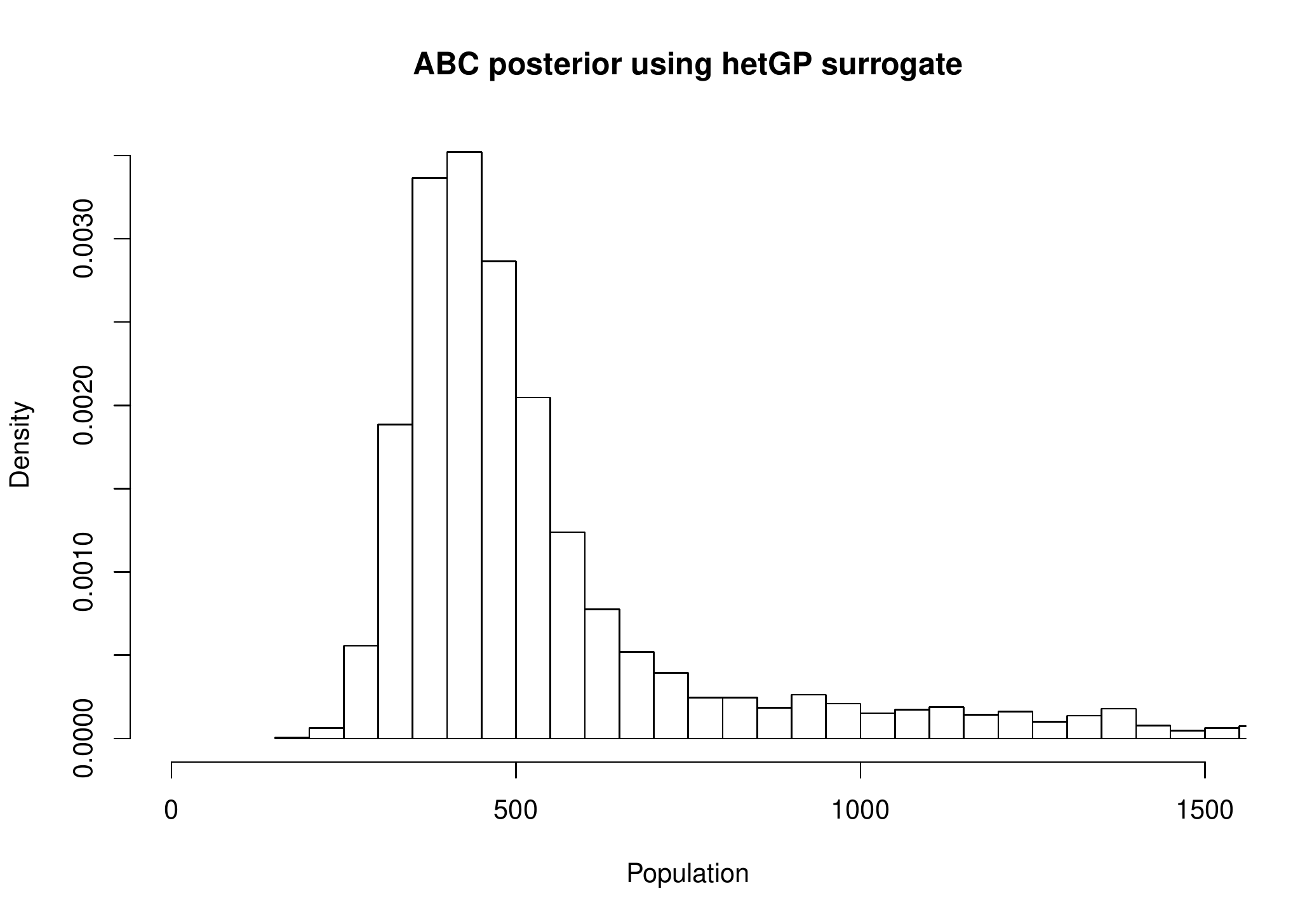}
\caption{ABC fish calibration: directly (left) and via hetGP surrogate (right). The prior for the true population size is uniform on $\{200,\dots,4000\}$, and 25 recaptured fish were observed. The left plot shows 10,000 simulations, highlights the 52 in agreement with the observation, and the histogram of accepted simulations. The histogram in the right panel is for the 3811 accepted draws out of 1,000,000 from hetGP.}
\label{fig:FishABC}
\end{figure}

Here we apply ABC to the fish simulator in order to estimate how many fish are in the population.   Suppose that 25 fish are recaptured in the second round. A straightforward method to determine the total population size is to simulate many times from the NetLogo fish model, for many different values of the total fish population, and ``accept'' a simulation every time it leads to 25 fish being recaptured. This is exactly ABC, and is a fairly common practice with ABMs. Doing so 10,000 times, using a uniform prior on the integers between 200 and 4000, so each such population size has prior probability 1/3801, yields the results in the left panel of Figure \ref{fig:FishABC}. This direct use of the simulator produces only 52 accepted samples, which is a very small number, and this from 10,000 simulator runs. In comparison, a hetGP surrogate fit from only 400 runs, and from which 1,000,000 samples can be quickly drawn, yields 3811 accepted samples. This result, illustrated in the right panel of Figure \ref{fig:FishABC}, gives a less noisy histogram with the same overall shape. If the agent-based model is even marginally costly then a surrogate is unquestionably valuable for ABC computations.
 
\subsection{Related Calibration Techniques} 
\label{sec:OtherCalib}


Bound-to-Bound \citep{frenklach2016comparison} is akin to HM, where an error bound that  sweeps up all uncertainties is similarly defined and quadratic programming is then used to find feasible bounds for $u_C$.  Bayesian Melding \citep{poole2000inference, raftery1995inference} is a technique related to Bayesian calibration, used to reconcile differences between elicited prior distributions on inputs and outputs of a simulator. It has been applied in ecology, epidemiology, urban modeling, and pollution monitoring  \citep{vsevvcikova2007assessing,alkema2007probabilistic,radtke2002bayesian,fuentes2005model}.

\section{Other Methods and Objectives}
\label{sec:other}

Here we briefly outline other surrogate modeling and downstream tasks.

\subsection{Regression Trees}
\label{sec:cattree}

In some situations the simulator mean $M$ may have discontinuities or “regime changes”, where a very different relationship between $y$ and $x$ exists in one part of input space compared to another part (i.e., non-stationarity). Regression Trees \citep{breiman1984classification} form a class of methods that can be useful in these situations. They are also useful in contexts where some inputs are categorical rather than numerical. The problems are treated by dividing the input space into mutually exclusive regions within which independent surrogates (GPs or other regression methods) are fit. 
 
Two approaches: the treed GP \citep[TGP][]{gramacy2008bayesian} and Bayesian Additive Regression Trees \citep[BART][]{chipman2010bart} have found wide application. Both use the data to automatically partition the input space, rely on Bayesian computation, and have public software: TGP in {\tt tgp} on CRAN \citep{tgp,gramacy2007tgp}; BART in several {\sf R} packages, including {\tt BayesTree} \citep{BayesTree} and {\tt BART} \citep{BART}.  
 
Other approaches by \cite{rulliere2018nested}, and via Voronoi tessellations instead of trees \citep[e.g.,][]{kim2005analyzing,rushdi2017vps,park2018patchwork}, have received less attention. \cite{pratola2020heteroscedastic} extends BART to heteroscedastic $\sigma_v^2$ (HBART) by modeling $M$ as a sum of Bayesian regression trees (as in BART) and the intrinsic variance $\sigma^2_v(x)$ as a product of Bayesian regression trees, in a joint approach similar to that in Section \ref{sec:hetGP}.

Calibration methods capitalizing on the KOH approach and using regression trees as in TGP (Section \ref{sec:cattree}) are explored in \cite{konomi2017bayesian}. In each terminal node of the partition a GP with an independent constant intrinsic variance term is assumed for the computer model output.  An independent GP is also deployed for the discrepancy term. Though $\sigma_v^2$ is constant at each terminal node the constants can vary across the terminal nodes so heteroscedasticity is automatically incorporated.

\subsection{Qualitative Inputs}
\label{sec:QualIn}
Categorical (qualitative) variables are often present in stochastic simulators, especially those that incorporate characteristics of human behavior. While regression trees  are capable of dealing with categorical inputs \citep{broderick2011classification,gramacy2010categorical}, GPs may be more effective as surrogates for smooth simulator output.

\citet{qian2008gaussian}, \citet{zhou2011simple}, and \citet{chen2013stochastic} describe ways to extend the kernels used for numerical inputs to incorporate qualitative variables. Painting with a broad brush, their approaches take the correlation between two outputs $y(x_i)$ and $y(x_j)$ as the product of two correlation functions: $C_c(w_i, w_j)$ dealing with the continuous inputs, $w$, and $C_q(z_i, z_j)$ for the qualitative variables, $z$. A simple way of building $C_q$ takes
\begin{equation} \label{eq:qualgeneral}
C_q(w_i, w_j) = \prod_{k=1}^{K} \tau_{k, w_{ik},w_{jk}} 
\end{equation}
where $K$ is the number of qualitative variables and $\tau_{k, w_{ik},w_{jk}}$ represents the correlation between $w_{ik}$ and $w_{jk}$. One example of $\tau_{k, w_{ik},w_{jk}}$ is:
\begin{equation} \label{eq:multiqualpar}
    \tau_{j, w_{ik},w_{jk}} = \exp\{-(\phi_{ik} + \phi_{jk})I[w_{ik} \neq w_{jk}]\}
\end{equation}
where $I$ is the indicator function 
(= 1 if its argument is true, = 0 if false),
and $\phi > 0$. The cited references also provide other ways of modeling $\tau_{k, w_{ik},w_{jk}}$.
Alternative methods exist e.g., \cite{Zhang2019latent} make use of latent variables for qualitative models.

\subsection{Optimization}
\label{sec:opt}

A common experimental objective is to maximize an output of the simulator, i.e., to find an input $x_{\mathrm{max}}$ that maximizes the output $y(x)$. For minimization instead, replace $y(x)$ by $-y(x)$. Optimisation is usually a sequential process where successive $x$s are chosen to get closer and closer to the optimal $x_{\mathrm{max}}$ — a sequential design problem (see Section \ref{sec:design}). With stochastic simulators, $y(x)$ is random, and optima are less concretely defined — the output is different every time the simulator is run at the same $x$. As a consequence, interest usually lies in maximizing a non-random quantity of interest, such as the mean, $M$, or possibly another scalar quantity such as the $q^{th}$ quantile.

For deterministic simulators Bayesian optimization \citep{Mockus1978, Jones1998} is a popular technique.  An initial set of runs is used to build a GP surrogate and new runs are chosen by maximizing an ``acquisition function'' $\alpha(x)$.  Iteratively choosing $x_{\mathrm{new}} = \arg\max_{x} \alpha(x)$) provides a progressively improved estimate for the maximum.  A widely used choice for $\alpha(x)$ is the {\em expected improvement} (EI):
\begin{equation}
\label{eq:EI}
   \alpha_{\mathrm{EI}}(x) = E[ \max \left(y(x) - y_{\mathrm{max}}, 0\right)]. 
\end{equation}
Maximizing EI chooses the input $x_{\mathrm{new}}$ that maximizes the expected increase in the maximum value, $y_{\mathrm{max}}$,  of already observed runs. With $y$ modeled by a GP:
\begin{equation}
\label{eq:EIGP}
         \alpha_{\mathrm{EI}}(x) = (y_{\mathrm{max}} - \mu_N(x)) \Phi\left(\frac{\mu_N(x) - y_{\mathrm{max}}}{\sigma_N(x)}\right) + \sigma_N(x) \phi\left(\frac {y_{\mathrm{max}} - \mu_N(x)}{\sigma_N(x)}\right)
\end{equation}
where $\mu_N(x)$ is the predictive mean of the GP, $\sigma_N(x)$ its standard deviation, $\phi$ is the standard normal density, and $\Phi$ the standard normal distribution function. 

Alternative acquisition functions have generated extensive work on Bayesian optimization in recent years, mostly in the machine learning literature. The probability of improvement \citep{kushner1964new} is an early example, and others, such as the GP upper confidence bound (GP-UCB) \citep{srinivas2009gaussian}, consider homoscedastic simulator error. A recent summary can be found in \cite{Frazier2018}.”

For stochastic simulators, the EI procedure can be extended by replacing $y_{\mathrm{max}}$, now a random variable, with the maximum estimated mean of currently run inputs, $\mu_{\mathrm{max}} = \max_{i \in \{1,\dots,N\}} \mu_N(x_i)$, see \citet{Vazquez2008}.  Or, one can seek improvement over the maximum estimated mean of any possible input, $\max_{x}\mu_N(x)$ \citep{gramacy:lee:2011}. In these cases, the $\sigma_N(x)$ term must exclude the $\sigma_v^2(x)$ term that comes from say, a hetGP. Implementation of this method is provided in the {\tt hetGP} package.

Alternative criteria for stochastic problems with constant intrinsic noise are discussed and compared in \cite{picheny2013benchmark}; with the above method is referred to as the ``plugin'' method.  An  {\sf R} package for implementing several of these choices is available in {\tt DiceOptim} (\cite{DiceOptim, picheny2014noisy}). \cite{Jalali2017comparison} also do a similar comparison for heteroscedastic noise.

The related goal of level set estimation to find regions where the output exceeds a threshold $T$ can also be targeted with sequential criteria similar to EI. A simple criterion is maximum contour uncertainty (MCU), wherein new points are chosen according to a weighted sum of how close to $T$ they are believed to be and the degree of uncertainty for that point. \cite{lyuetal2018} provide some discussion here. This method is also implemented in {\tt hetGP}.

Optimization using Gaussian processes, specifically in the presence of intrinsic variability that is potentially heteroscedastic (and potentially non-normal) is an interesting research question and possibly deserving of its own review. Nonetheless, the references provided here should provide a good introduction.

\subsection{Sensitivity Analysis}
\label{sec:sensitivity}

Determining and measuring the effect of inputs on the output is usually part of any simulator experiment. Doing so assists scientific understanding of the system and enables screening out potentially superfluous variables.  This goal has many related names: sensitivity analysis, screening, variable selection, etc., but the overall objective is generally the same $-$ summarize and measure the influence of each input.

For deterministic simulators, Sobol indices \citep{sobol1993sensitivity} are widely used.
Probabilistic distributions are assumed on the inputs of the simulator in order to represent their range of variations. Then, a functional Analysis of Variance (ANOVA) decomposition splits the variation of the simulator output into multiple components, each representing the individual contribution of an input variable $x_j$ or combination of input variables. A Sobol index is then computed as the percentage of the total simulator output variation explained by a component. Key Sobol indices include main effects (the percentage of variation explained by the individual $x_j$s \emph{alone}) and variation explained by interactive additive effects with other inputs. 
Computing the components takes large numbers of runs but the use of surrogate GPs make the calculations feasible \citep{schonlau2006screening, marrel2009calculations}. An enveloping discussion of sensitivity is provided by \cite{oakley2004probabilistic}.

Two extensions, by \citet{Marrel2012global} and \citet{hart2017efficient}, of  Sobol indices for stochastic simulators yield the following expression for the stochastic simulator:
\begin{equation}
\label{eq:StoformSA}
y(x) = y(x,\epsilon_{\mathrm{seed}}) 
\end{equation}
where $x$ is the set of controllable inputs.  The input $\epsilon_{\mathrm{seed}}$ is responsible for output stochasticity, standing in for intrinsic variability, and is sometimes called a seed variable. As with a deterministic simulator, a probabilistic distribution (typically uniform) is assumed to represent the range of variation in controllable inputs.

In \cite{Marrel2012global}, the total variation in the \emph{mean} of the stochastic simulator is analysed through a functional ANOVA decomposition and Sobol indices are computed based on the percentage of the total \emph{simulator} variation each component explains. The variation explained by the seed variable $\epsilon_{\mathrm{seed}}$ can also be computed, representing the total variation explained by the intrinsic variance. Additionally, a sensitivity analysis of the intrinsic variance $\sigma^2_N(x)$ can be conducted separately to gather information on which input variables most impact the heteroscedasticity.

\cite{hart2017efficient} assumes the simulator can be run at different inputs $x$ with the same seed $\epsilon_{\mathrm{seed}}$. Rather than building a joint stochastic simulator surrogate for the mean and variance, as described in Section \ref{sec:hetGP}, they build a separate surrogate for a number of seeds. For each seed, they obtain a realization of each Sobol index, and by aggregating the realizations, they obtain distributions for the indices. 

The extensive literature on model selection may have counterparts that can be effective for stochastic simulators. But a fully satisfactory approach even for deterministic simulators remains somewhat elusive. 

\section{Concluding Remarks}
\label{sec:conclusions}

There are several key messages to be drawn from this review, each pointing to open or new research questions:

\textbf{Gaussian Process Surrogates.} GPs are discussed extensively  because they provide a flexible way of allowing the data to inform about the shape of the underlying process. Moreover, they can be effective predictors and quantifiers of uncertainty. Diagnosing shortcomings in a GP for stochastic simulators (available in deterministic settings \citep{bastos2009diagnostics}) is not yet well-established.

As noted in Section \ref{sec:models}, neural network (deep learning) methods are in active use and under study, some of which may, in combination with GPs, offer promising research directions \citep{schultz2018practical}.

Additionally, it can be difficult to effectively capture non-normal variability. Doing so with as few simulations as possible, whilst also properly quantifying the various uncertainties, is likely to be an important research direction for stochastic simulator analysis. The wider quantile regression literature is likely a good starting point.

\textbf{Design.} Stochastic simulators differ from deterministic ones because they require much larger sample sizes and permit the use of replicates, whose treatment is generally ad hoc. This leads to the questions raised in Section \ref{sec:design}, forming a direction of important research. Design size rules of thumb, useful even if imperfect, exist for deterministic simulators \citep{Loeppky2009}, but are lacking for stochastic simulators. 

\textbf{Calibration.}  Accounting for model discrepancy in calibration is critical but there is no obvious “one-size-fits-all” method.  A broad empirical comparison is needed with guidance about which strategies are effective under which conditions. Assessing the effectiveness of different methods can be challenging \citep[see][for one comparison between ABC and HM]{Mckinley2018approximate}, but sorely needed. 

\textbf{Simulator Complexity}  For complex stochastic simulators it may not be feasible to obtain enough runs.  In some instances, the simulator can be replaced with a less complex one \citep[e.g.][]{molina2005statistical} that captures key features and permits adequate numbers of simulations. Another path, coupling stochastic simulators with deterministic simulators has been explored \citep{baker2020predicting} as a way to deal with low simulation budgets. Multi-fidelity modeling, where multiple simulators of varying complexity are coupled together \citep{kennedy2000predicting, kennedy2020multilevel} is a promising solution where possible.

In a similar vein, certain outputs may be less noisy than others, and the modeling of the less-noisy outputs can improve the modeling of the noisier ones.  For example \citet{wang2020enhancing} use the expectation of a simulator to improve the estimation of noisier quantiles. This is related to the wider variance reduction literature, which has a long history \citep{barton2017history}. Variance reduction has been applied in a number of examples but its use in ABMs is not apparent, perhaps due to the profusion of stochastic elements in an ABM. 
Fixing the initial seed in a stochastic simulator has played a role in sensitivity analysis (see Section \ref{sec:sensitivity}), but leveraging information about the intrinsic randomness for wider purposes is an open problem. 
\\

This review strives to raise awareness of existing tools and strategies for treating stochastic simulators and provide a starting point for practitioners interested in utilizing up-to-date statistical approaches. Despite the problems being pervasive and challenging there is a shortage of statistical research in this field. The problems pose computational and technical questions, as well as theoretical and philosophical ones. Current solutions are often capable, but there is a lack of comprehensive comparison between different solutions, and a lack of testing regarding their generalizability to complex situations (such as very large data sets). The hope is that the review provokes statistical researchers to engage the open questions discussed.

\subsubsection*{Acknowledgements}

We gratefully acknowledge the support and funding provided by SAMSI during the Model Uncertainty: Mathematical and Statistical program, 2018-19. Moreover,  Professor David Banks, Director of SAMSI, is thanked for suggesting and encouraging the  writing of this review. 

Robert Gramacy was supported in part by  DOE LAB 17-1697 via a subaward from
Argonne National Laboratory for SciDAC/DOE Office of Science ASCR and High
Energy Physics, and by the National Science Foundation DMS-1821258.

Pierre Barbillon has received support from the Marie-Curie FP7 COFUND People Programme of the European Union, through the award of an AgreenSkills/AgreenSkills+ fellowship (under grant agreement n$^{\circ}$609398).

\appendix
\section{Appendix - Ocean Truth}
\label{sec:appendix_truth}
Throughout, reference to plots of the ``truth'' of the Ocean model is made. These plots are presented here, as well as in the supplementary material, for convenience.

\begin{figure}[hbt!]
\centering
\includegraphics[scale=0.45,trim=5 0 5 0]{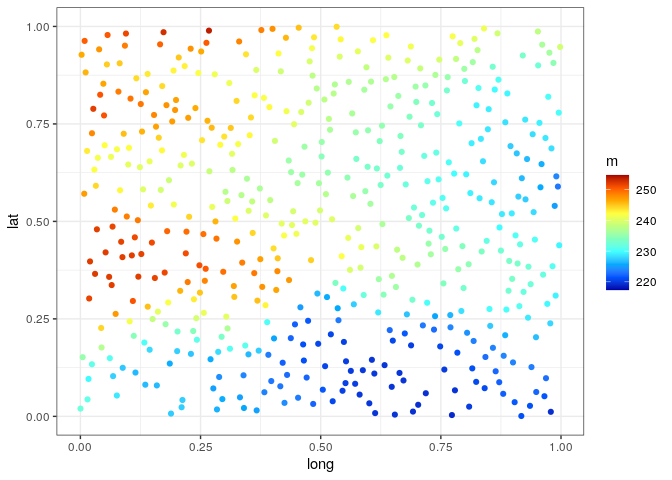}
\includegraphics[scale=0.45,trim=40 0 5 0,clip=TRUE]{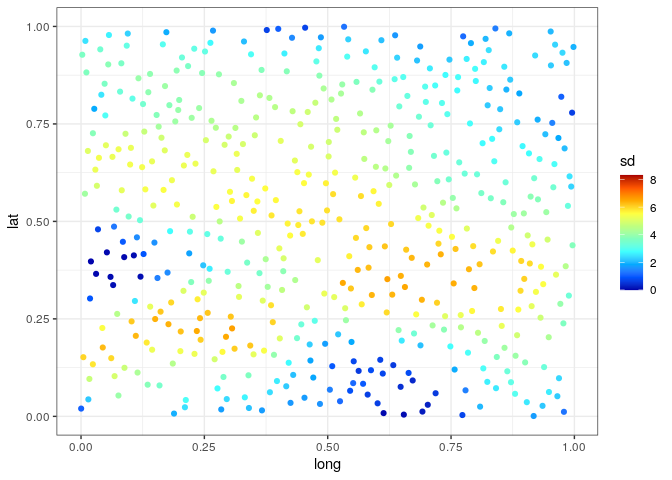}
\caption{The ``true'' mean and standard deviation for the Ocean model, for 500 different sites.}
\label{fig:OceanTruth}
\end{figure}

\bibliographystyle{apalike}
\bibliography{references}

\begin{thebibliography}{}

\bibitem[Abt and Welch, 1998]{abt1998fisher}
Abt, M. and Welch, W.~J. (1998).
\newblock Fisher information and maximum-likelihood estimation of covariance
  parameters in {G}aussian stochastic processes.
\newblock {\em Canadian Journal of Statistics}, 26(1):127--137.

\bibitem[Alkema et~al., 2007]{alkema2007probabilistic}
Alkema, L., Raftery, A.~E., Clark, S.~J., et~al. (2007).
\newblock Probabilistic projections of {{HIV}} prevalence using {B}ayesian
  melding.
\newblock {\em The Annals of Applied Statistics}, 1(1):229--248.

\bibitem[Andrianakis et~al., 2017]{Andrianakis2017efficient}
Andrianakis, I., McCreesh, N., Vernon, I., McKinley, T.~J., Oakley, J.~E.,
  Nsubuga, R.~N., Goldstein, M., and White, R.~G. (2017).
\newblock {Efficient history matching of a high dimensional individual-based
  {HIV} transmission model}.
\newblock {\em SIAM/ASA Journal on Uncertainty Quantification}, 5(1):694--719.

\bibitem[Andrianakis et~al., 2015]{Andrianakis2015bayesian}
Andrianakis, I., Vernon, I.~R., McCreesh, N., McKinley, T.~J., Oakley, J.~E.,
  Nsubuga, R.~N., Goldstein, M., and White, R.~G. (2015).
\newblock {B}ayesian history matching of complex infectious disease models
  using emulation: a tutorial and a case study on {HIV} in {U}ganda.
\newblock {\em PLoS Computational Biology}, 11(1):e1003968.

\bibitem[Ankenman et~al., 2010]{Ankenman2010}
Ankenman, B., Nelson, B.~L., and Staum, J. (2010).
\newblock Stochastic kriging for simulation metamodeling.
\newblock {\em Operations Research}, 58(2):371--382.

\bibitem[Ba, 2019]{SLHD}
Ba, S. (2019).
\newblock {\em SLHD: Maximin-Distance (Sliced) Latin Hypercube Designs}.
\newblock R package version 1.1.2.

\bibitem[Ba et~al., 2012]{ba2012composite}
Ba, S., Joseph, V.~R., et~al. (2012).
\newblock Composite {G}aussian process models for emulating expensive
  functions.
\newblock {\em The Annals of Applied Statistics}, 6(4):1838--1860.

\bibitem[Baker et~al., 2020]{baker2020predicting}
Baker, E., Challenor, P., and Eames, M. (2020).
\newblock Predicting the output from a stochastic computer model when a
  deterministic approximation is available.
\newblock {\em Journal of Computational and Graphical Statistics}, in press.

\bibitem[Barton et~al., 2017]{barton2017history}
Barton, R., Nakayama, M.~K., and Schruben, L. (2017).
\newblock History of improving statistical efficiency.
\newblock In {\em 2017 Winter Simulation Conference (WSC)}, pages 158--180.
  IEEE.

\bibitem[Bastos and O’Hagan, 2009]{bastos2009diagnostics}
Bastos, L.~S. and O’Hagan, A. (2009).
\newblock Diagnostics for {G}aussian process emulators.
\newblock {\em Technometrics}, 51(4):425--438.

\bibitem[Bayarri et~al., 2007a]{bayarri2007computer}
Bayarri, M., Berger, J., Cafeo, J., Garcia-Donato, G., Liu, F., Palomo, J.,
  Parthasarathy, R., Paulo, R., Sacks, J., Walsh, D., et~al. (2007a).
\newblock Computer model validation with functional output.
\newblock {\em The Annals of Statistics}, 35(5):1874--1906.

\bibitem[Bayarri et~al., 2009]{bayarri2009predicting}
Bayarri, M.~J., Berger, J.~O., Kennedy, M.~C., Kottas, A., Paulo, R., Sacks,
  J., Cafeo, J.~A., Lin, C.-H., and Tu, J. (2009).
\newblock Predicting vehicle crashworthiness: Validation of computer models for
  functional and hierarchical data.
\newblock {\em Journal of the American Statistical Association},
  104(487):929--943.

\bibitem[Bayarri et~al., 2007b]{bayarri2007}
Bayarri, M.~J., Berger, J.~O., Paulo, R., Sacks, J., Cafeo, J.~A., Cavendish,
  J., Lin, C.-H., and Tu, J. (2007b).
\newblock A framework for validation of computer models.
\newblock {\em Technometrics}, 49(2):138--154.

\bibitem[Begon et~al., 1979]{begon1979investigating}
Begon, M. et~al. (1979).
\newblock {\em Investigating Animal Abundance: Capture-Recapture for
  Biologists.}
\newblock Edward Arnold Publishers Ltd, London.

\bibitem[Bernardo et~al., 1992]{bernardo1992integrated}
Bernardo, M.~C., Buck, R., Liu, L., Nazaret, W.~A., Sacks, J., and Welch, W.~J.
  (1992).
\newblock Integrated circuit design optimization using a sequential strategy.
\newblock {\em IEEE Transactions on Computer-Aided Design of Integrated
  Circuits and Systems}, 11(3):361--372.

\bibitem[Binois and Gramacy, 2018]{hetGP}
Binois, M. and Gramacy, R.~B. (2018).
\newblock {\em hetGP: Heteroskedastic {G}aussian Process Modeling and Design
  under Replication}.
\newblock R package version 1.1.1.

\bibitem[Binois et~al., 2018a]{Binois2018JCGS}
Binois, M., Gramacy, R.~B., and Ludkovski, M. (2018a).
\newblock Practical heteroscedastic {G}aussian process modeling for large
  simulation experiments.
\newblock {\em Journal of Computational and Graphical Statistics},
  27(4):808--821.

\bibitem[Binois et~al., 2018b]{Binois2018TECH}
Binois, M., Huang, J., Gramacy, R.~B., and Ludkovski, M. (2018b).
\newblock Replication or exploration? {S}equential design for stochastic
  simulation experiments.
\newblock {\em Technometrics}, 61(1):7--23.

\bibitem[Bisset et~al., 2009]{bisset2009epifast}
Bisset, K.~R., Chen, J., Feng, X., Kumar, V., and Marathe, M.~V. (2009).
\newblock Epifast: a fast algorithm for large scale realistic epidemic
  simulations on distributed memory systems.
\newblock In {\em Proceedings of the 23rd international conference on
  Supercomputing}, pages 430--439. ACM.

\bibitem[Boukouvalas and Cornford, 2009]{Boukouvalas2009Learning}
Boukouvalas, A. and Cornford, D. (2009).
\newblock Learning heteroscedastic {G}aussian processes for complex datasets.
\newblock Technical report, Aston University, Neural Computing Research Group.

\bibitem[Boukouvalas et~al., 2014a]{boukouvalas2014optimal}
Boukouvalas, A., Cornford, D., and Stehl{\'\i}k, M. (2014a).
\newblock Optimal design for correlated processes with input-dependent noise.
\newblock {\em Computational Statistics \& Data Analysis}, 71:1088--1102.

\bibitem[Boukouvalas et~al., 2014b]{Boukouvalas2014Calibration}
Boukouvalas, A., Sykes, P., Cornford, D., and Maruri-Aguilar, H. (2014b).
\newblock {B}ayesian precalibration of a large stochastic microsimulation
  model.
\newblock {\em IEEE Transactions on Intelligent Transportation Systems},
  15(3):1337--1347.

\bibitem[Breiman et~al., 1984]{breiman1984classification}
Breiman, L., Friedman, J., Olshen, R., and Stone, C. (1984).
\newblock {C}lassification and {R}egression {T}rees.

\bibitem[Broderick and Gramacy, 2011]{broderick2011classification}
Broderick, T. and Gramacy, R. (2011).
\newblock Classification and categorical inputs with treed {G}aussian process
  models.
\newblock {\em Journal of Classification}, 28(2):244--270.

\bibitem[Brynjarsd{\'o}ttir and O'Hagan, 2014]{bryn2014learning}
Brynjarsd{\'o}ttir, J. and O'Hagan, A. (2014).
\newblock Learning about physical parameters: The importance of model
  discrepancy.
\newblock {\em Inverse Problems}, 30(11):114007.

\bibitem[Chen et~al., 2016]{chen2016analysis}
Chen, H., Loeppky, J.~L., Sacks, J., Welch, W.~J., et~al. (2016).
\newblock Analysis methods for computer experiments: how to assess and what
  counts?
\newblock {\em Statistical Science}, 31(1):40--60.

\bibitem[Chen et~al., 2013]{chen2013stochastic}
Chen, X., Wang, K., and Yang, F. (2013).
\newblock Stochastic kriging with qualitative factors.
\newblock In {\em 2013 Winter Simulations Conference (WSC)}, pages 790--801.
  IEEE.

\bibitem[Chipman et~al., 2010]{chipman2010bart}
Chipman, H., George, E., and McCulloch, R. (2010).
\newblock {BART}: {B}ayesian additive regression trees.
\newblock {\em The Annals of Applied Statistics}, 4(1):266--298.

\bibitem[Chipman and McCulloch, 2016]{BayesTree}
Chipman, H. and McCulloch, R. (2016).
\newblock {\em BayesTree: {B}ayesian Additive Regression Trees}.
\newblock R package version 0.3-1.4.

\bibitem[Chung et~al., 2019]{chung2019parameter}
Chung, M., Binois, M., Gramacy, R.~B., Bardsley, J.~M., Moquin, D.~J., Smith,
  A.~P., and Smith, A.~M. (2019).
\newblock Parameter and uncertainty estimation for dynamical systems using
  surrogate stochastic processes.
\newblock {\em SIAM Journal on Scientific Computing}, 41(4):A2212--A2238.

\bibitem[Conti and O’Hagan, 2010]{conti2010}
Conti, S. and O’Hagan, A. (2010).
\newblock {B}ayesian emulation of complex multi-output and dynamic computer
  models.
\newblock {\em Journal of Statistical Planning and Inference}, 140(3):640 --
  651.

\bibitem[Craig et~al., 1997]{craig1997pressure}
Craig, P.~S., Goldstein, M., Seheult, A.~H., and Smith, J.~A. (1997).
\newblock Pressure matching for hydrocarbon reservoirs: a case study in the use
  of bayes linear strategies for large computer experiments.
\newblock In {\em Case Studies in {B}ayesian Statistics}, pages 37--93.
  Springer.

\bibitem[Currin et~al., 1991]{currin1991bayesian}
Currin, C., Mitchell, T., Morris, M., and Ylvisaker, D. (1991).
\newblock {B}ayesian prediction of deterministic functions, with applications
  to the design and analysis of computer experiments.
\newblock {\em Journal of the American Statistical Association},
  86(416):953--963.

\bibitem[Damblin et~al., 2018]{damblin2018adaptive}
Damblin, G., Barbillon, P., Keller, M., Pasanisi, A., and Parent, {\'E}.
  (2018).
\newblock Adaptive numerical designs for the calibration of computer codes.
\newblock {\em SIAM/ASA Journal on Uncertainty Quantification}, 6(1):151--179.

\bibitem[Duan et~al., 2017]{Duan:2017}
Duan, W., Ankenman, B.~E., Sanchez, S.~M., and Sanchez, P.~J. (2017).
\newblock Sliced full factorial-based latin hypercube designs as a framework
  for a batch sequential design algorithm.
\newblock {\em Technometrics}, 59(1):11--22.

\bibitem[Erickson et~al., 2018]{Erickson:2018}
Erickson, C.~B., Ankenman, B.~E., Plumlee, M., and Sanchez, S.~M. (2018).
\newblock Gradient based criteria for sequential design.
\newblock In {\em 2018 Winter Simulation Conference (WSC)}, pages 467--478.
  IEEE.

\bibitem[Fadikar et~al., 2018]{fadikar2018}
Fadikar, A., Higdon, D., Chen, J., Lewis, B., Venkatramanan, S., and Marathe,
  M. (2018).
\newblock Calibrating a stochastic, agent-based model using quantile-based
  emulation.
\newblock {\em SIAM/ASA Journal on Uncertainty Quantification},
  6(4):1685--1706.

\bibitem[Farah et~al., 2014]{Farah2014bayesian}
Farah, M., Birrell, P., Conti, S., and Angelis, D.~D. (2014).
\newblock {B}ayesian emulation and calibration of a dynamic epidemic model for
  {A/H1N1} influenza.
\newblock {\em Journal of the American Statistical Association},
  109(508):1398--1411.

\bibitem[Feynman, 1948]{feynman1948}
Feynman, R. (1948).
\newblock Space-time approach to non-relativistic quantum mechanics.
\newblock {\em Reviews of Modern Physics}, 20(2):367--387.

\bibitem[Frazier, 2018]{Frazier2018}
Frazier, P.~I. (2018).
\newblock A tutorial on {B}ayesian optimization.
\newblock {\em arXiv preprint arXiv:1807.02811}.

\bibitem[Frenklach et~al., 2016]{frenklach2016comparison}
Frenklach, M., Packard, A., Garcia-Donato, G., Paulo, R., and Sacks, J. (2016).
\newblock Comparison of statistical and deterministic frameworks of uncertainty
  quantification.
\newblock {\em SIAM/ASA Journal on Uncertainty Quantification}, 4(1):875--901.

\bibitem[Fricker et~al., 2013]{fricker2013}
Fricker, T.~E., Oakley, J.~E., and Urban, N.~M. (2013).
\newblock Multivariate {G}aussian process emulators with nonseparable
  covariance structures.
\newblock {\em Technometrics}, 55(1):47--56.

\bibitem[Fuentes and Raftery, 2005]{fuentes2005model}
Fuentes, M. and Raftery, A.~E. (2005).
\newblock Model evaluation and spatial interpolation by {B}ayesian combination
  of observations with outputs from numerical models.
\newblock {\em Biometrics}, 61(1):36--45.

\bibitem[Gal and Ghahramani, 2016]{gal2016dropout}
Gal, Y. and Ghahramani, Z. (2016).
\newblock Dropout as a {B}ayesian approximation: representing model uncertainty
  in deep learning.
\newblock In {\em Proceedings of the 33rd International Conference on
  International Conference on Machine Learning}, pages 1050--1059.

\bibitem[Gao et~al., 1996]{gao1996predicting}
Gao, F., Sacks, J., and Welch, W.~J. (1996).
\newblock Predicting urban ozone levels and trends with semiparametric
  modeling.
\newblock {\em Journal of Agricultural, Biological, and Environmental
  Statistics}, 1(4):404--425.

\bibitem[Gneiting and Raftery, 2007]{gneiting2007strictly}
Gneiting, T. and Raftery, A.~E. (2007).
\newblock Strictly proper scoring rules, prediction, and estimation.
\newblock {\em Journal of the American Statistical Association},
  102(477):359--378.

\bibitem[Goldberg et~al., 1997]{Goldberg1998regression}
Goldberg, P.~W., Williams, C.~K., and Bishop, C.~M. (1997).
\newblock Regression with input-dependent noise: a {G}aussian process
  treatment.
\newblock In {\em Proceedings of the 10th International Conference on Neural
  Information Processing Systems}, pages 493--499.

\bibitem[Gramacy, 2007]{gramacy2007tgp}
Gramacy, R. (2007).
\newblock {tgp}: an {R} package for {B}ayesian nonstationary, semiparametric
  nonlinear regression and design by treed {G}aussian process models.
\newblock {\em Journal of Statistical Software}, 19(9):6.

\bibitem[Gramacy and Lee, 2008]{gramacy2008bayesian}
Gramacy, R. and Lee, H. (2008).
\newblock {B}ayesian treed {G}aussian process models with an application to
  computer modeling.
\newblock {\em Journal of the American Statistical Association},
  103(483):1119--1130.

\bibitem[Gramacy and Taddy, 2010]{gramacy2010categorical}
Gramacy, R. and Taddy, M. (2010).
\newblock Categorical inputs, sensitivity analysis, optimization and importance
  tempering with tgp version 2, an {R} package for treed {G}aussian process
  models.
\newblock {\em Journal of Statistical Software}, 33(6):1--48.

\bibitem[Gramacy, 2020]{gramacy2020surrogates}
Gramacy, R.~B. (2020).
\newblock {\em Surrogates: {G}aussian Process Modeling, Design and Optimization
  for the Applied Sciences}.
\newblock Chapman Hall/CRC, Boca Raton, Florida.
\newblock \url{http://bobby.gramacy.com/surrogates/}.

\bibitem[Gramacy and Lee, 2009]{gra:lee:2009}
Gramacy, R.~B. and Lee, H. K.~H. (2009).
\newblock Adaptive design and analysis of supercomputer experiments.
\newblock {\em Technometrics}, 51(2):130--145.

\bibitem[Gramacy and Lee, 2011]{gramacy:lee:2011}
Gramacy, R.~B. and Lee, H. K.~H. (2011).
\newblock Optimization under unknown constraints.
\newblock In Bernardo, J., Bayarri, S., Berger, J.~O., Dawid, A.~P., Heckerman,
  D., Smith, A. F.~M., and West, M., editors, {\em {B}ayesian Statistics 9},
  pages 229--256. Oxford University Press.

\bibitem[Gramacy and Taddy, 2016]{tgp}
Gramacy, R.~B. and Taddy, M.~A. (2016).
\newblock {\em tgp: {B}ayesian Treed {G}aussian Process Models}.
\newblock R package version 2.4-14.

\bibitem[Graves, 2011]{graves2011practical}
Graves, A. (2011).
\newblock Practical variational inference for neural networks.
\newblock In {\em Proceedings of the 24th International Conference on Neural
  Information Processing Systems}, pages 2348--2356.

\bibitem[Grimm et~al., 2006]{grimm2006standard}
Grimm, V., Berger, U., Bastiansen, F., Eliassen, S., Ginot, V., Giske, J.,
  Goss-Custard, J., Grand, T., Heinz, S.~K., Huse, G., et~al. (2006).
\newblock A standard protocol for describing individual-based and agent-based
  models.
\newblock {\em Ecological Modelling}, 198(1-2):115--126.

\bibitem[Gu and Wang, 2018]{gu2018scaled}
Gu, M. and Wang, L. (2018).
\newblock Scaled {G}aussian stochastic process for computer model calibration
  and prediction.
\newblock {\em SIAM/ASA Journal on Uncertainty Quantification},
  6(4):1555--1583.

\bibitem[Hart et~al., 2017]{hart2017efficient}
Hart, J.~L., Alexanderian, A., and Gremaud, P.~A. (2017).
\newblock Efficient computation of indices for stochastic models.
\newblock {\em SIAM Journal on Scientific Computing}, 39(4):A1514--A1530.

\bibitem[Harville, 1998]{harville1998matrix}
Harville, D.~A. (1998).
\newblock {\em Matrix algebra from a statistician's perspective}.
\newblock Springer-Verlag, New York.

\bibitem[Henderson et~al., 2009]{henderson2009bayesian}
Henderson, D.~A., Boys, R.~J., Krishnan, K.~J., Lawless, C., and Wilkinson,
  D.~J. (2009).
\newblock {B}ayesian emulation and calibration of a stochastic computer model
  of mitochondrial {DNA} deletions in substantia nigra neurons.
\newblock {\em Journal of the American Statistical Association},
  104(485):76--87.

\bibitem[Herbei and Berliner, 2014]{Herbei2014estimating}
Herbei, R. and Berliner, L.~M. (2014).
\newblock Estimating ocean circulation: An {MCMC} approach with approximated
  likelihoods via the bernoulli factory.
\newblock {\em Journal of the American Statistical Association},
  109(507):944--954.

\bibitem[Higdon et~al., 2008]{higdon2008computer}
Higdon, D., Gattiker, J., Williams, B., and Rightley, M. (2008).
\newblock Computer model calibration using high-dimensional output.
\newblock {\em Journal of the American Statistical Association},
  103(482):570--583.

\bibitem[Higdon et~al., 2004]{higdon2004combining}
Higdon, D., Kennedy, M., Cavendish, J.~C., Cafeo, J.~A., and Ryne, R.~D.
  (2004).
\newblock Combining field data and computer simulations for calibration and
  prediction.
\newblock {\em SIAM Journal on Scientific Computing}, 26(2):448--466.

\bibitem[Jalali et~al., 2017]{Jalali2017comparison}
Jalali, H., Van~Nieuwenhuyse, I., and Picheny, V. (2017).
\newblock Comparison of kriging-based methods for simulation optimization with
  heterogeneous noise.
\newblock {\em European Journal of Operational Research}, 261(1):279--301.

\bibitem[Johnson, 2010]{johnson2010implications}
Johnson, L.~R. (2010).
\newblock Implications of dispersal and life history strategies for the
  persistence of linyphiid spider populations.
\newblock {\em Ecological Modelling}, 221(8):1138--1147.

\bibitem[Johnson and Briggs, 2011]{johnson2011parameter}
Johnson, L.~R. and Briggs, C.~J. (2011).
\newblock Parameter inference for an individual based model of chytridiomycosis
  in frogs.
\newblock {\em Journal of Theoretical Biology}, 277(1):90--98.

\bibitem[Johnson et~al., 2018]{johnson2018phenomenological}
Johnson, L.~R., Gramacy, R.~B., Cohen, J., Mordecai, E., Murdock, C., Rohr, J.,
  Ryan, S.~J., Stewart-Ibarra, A.~M., Weikel, D., et~al. (2018).
\newblock Phenomenological forecasting of disease incidence using
  heteroskedastic {G}aussian processes: a dengue case study.
\newblock {\em The Annals of Applied Statistics}, 12(1):27--66.

\bibitem[Jolliffe, 2011]{jolliffe2011principal}
Jolliffe, I. (2011).
\newblock {\em Principal Component Analysis}.
\newblock Springer.

\bibitem[Jones et~al., 1998]{Jones1998}
Jones, D., Schonlau, M., and Welch, W. (1998).
\newblock Efficient global optimization of expensive black-box functions.
\newblock {\em Journal of Global Optimization}, 13(4):455--492.

\bibitem[Kac, 1949]{kac1949}
Kac, M. (1949).
\newblock On distributions of certain {W}iener functionals.
\newblock {\em Transactions of the American Mathematical Society}, 65:1--13.

\bibitem[Kennedy et~al., 2020]{kennedy2020multilevel}
Kennedy, J.~C., Henderson, D.~A., and Wilson, K.~J. (2020).
\newblock Multilevel emulation for stochastic computer models with an
  application to large offshore windfarms.
\newblock {\em arXiv preprint arXiv:2003.08921}.

\bibitem[Kennedy and O'Hagan, 2000]{kennedy2000predicting}
Kennedy, M.~C. and O'Hagan, A. (2000).
\newblock Predicting the output from a complex computer code when fast
  approximations are available.
\newblock {\em Biometrika}, 87(1):1--13.

\bibitem[Kennedy and O'Hagan, 2001]{kennedy2001bayesian}
Kennedy, M.~C. and O'Hagan, A. (2001).
\newblock {B}ayesian calibration of computer models.
\newblock {\em Journal of the Royal Statistical Society: Series B (Statistical
  Methodology)}, 63(3):425--464.

\bibitem[Kersaudy et~al., 2015]{kersaudy2015new}
Kersaudy, P., Sudret, B., Varsier, N., Picon, O., and Wiart, J. (2015).
\newblock A new surrogate modeling technique combining kriging and polynomial
  chaos expansions--application to uncertainty analysis in computational
  dosimetry.
\newblock {\em Journal of Computational Physics}, 286:103--117.

\bibitem[Kersting et~al., 2007]{Kersting2007MostLikely}
Kersting, K., Plagemann, C., Pfaff, P., and Burgard, W. (2007).
\newblock Most likely heteroscedastic {G}aussian process regression.
\newblock In {\em Proceedings of the 24th International Conference on Machine
  Learning}, pages 393--400.

\bibitem[Kim et~al., 2005]{kim2005analyzing}
Kim, H., Mallick, B., and Holmes, C. (2005).
\newblock Analyzing nonstationary spatial data using piecewise {G}aussian
  processes.
\newblock {\em Journal of the American Statistical Association},
  100(470):653--668.

\bibitem[Kleijnen, 2009]{kleijnen2009kriging}
Kleijnen, J.~P. (2009).
\newblock Kriging metamodeling in simulation: A review.
\newblock {\em European Journal of Operational Research}, 192(3):707--716.

\bibitem[Kleijnen, 2017]{kleijnen2017regression}
Kleijnen, J.~P. (2017).
\newblock Regression and kriging metamodels with their experimental designs in
  simulation: a review.
\newblock {\em European Journal of Operational Research}, 256(1):1--16.

\bibitem[Koenker and Bassett~Jr, 1978]{koenker1978regression}
Koenker, R. and Bassett~Jr, G. (1978).
\newblock Regression quantiles.
\newblock {\em Econometrica}, 46(1):33--50.

\bibitem[Konomi et~al., 2017]{konomi2017bayesian}
Konomi, B., Karagiannis, G., Lai, K., and Lin, G. (2017).
\newblock {B}ayesian treed calibration: an application to carbon capture with
  {AX} sorbent.
\newblock {\em Journal of the American Statistical Association},
  112(517):37--53.

\bibitem[Kushner, 1964]{kushner1964new}
Kushner, H.~J. (1964).
\newblock A new method of locating the maximum point of an arbitrary multipeak
  curve in the presence of noise.
\newblock {\em Journal of Basic Engineering}, 86(1):97--106.

\bibitem[Lakshminarayanan et~al., 2017]{lakshminarayanan2017simple}
Lakshminarayanan, B., Pritzel, A., and Blundell, C. (2017).
\newblock Simple and scalable predictive uncertainty estimation using deep
  ensembles.
\newblock In {\em Advances in Neural Information Processing Systems}, pages
  6402--6413.

\bibitem[Lee, 2015]{pyDOE}
Lee, A. (2015).
\newblock {\em pyDOE: The experimental design package for python}.
\newblock Python package version 0.3.8.

\bibitem[Liu et~al., 2009]{liu2009modularization}
Liu, F., Bayarri, M., and Berger, J. (2009).
\newblock Modularization in {B}ayesian analysis, with emphasis on analysis of
  computer models.
\newblock {\em {B}ayesian Analysis}, 4(1):119--150.

\bibitem[Loeppky et~al., 2009a]{Loeppky:2009}
Loeppky, J., Moore, L., and Williams, B. (2009a).
\newblock Batch sequential designs for computer experiments.
\newblock {\em Journal of Statistical Planning and Inference}, 140:1452--1464.

\bibitem[Loeppky et~al., 2009b]{Loeppky2009}
Loeppky, J.~L., Sacks, J., and Welch, W.~J. (2009b).
\newblock {Choosing the Sample Size of a Computer Experiment - A Practical
  Guide.}
\newblock {\em Technometrics}, 51(4):366--376.

\bibitem[Lyu et~al., 2018]{lyuetal2018}
Lyu, X., Binois, M., and Ludkovski, M. (2018).
\newblock Evaluating {G}aussian process metamodels and sequential designs for
  noisy level set estimation.
\newblock {\em arXiv preprint arXiv:1807.06712}.

\bibitem[Ma et~al., 2019]{ma2019computer}
Ma, P., Mondal, A., Konomi, B., Hobbs, J., Song, J., and Kang, E. (2019).
\newblock Computer model emulation with high-dimensional functional output in
  large-scale observing system uncertainty experiments.
\newblock {\em arXiv preprint arXiv:1911.09274}.

\bibitem[Marrel et~al., 2012]{Marrel2012global}
Marrel, A., Iooss, B., Da~Veiga, S., and Ribatet, M. (2012).
\newblock Global sensitivity analysis of stochastic computer models with joint
  metamodels.
\newblock {\em Statistics and Computing}, 22(3):833--847.

\bibitem[Marrel et~al., 2009]{marrel2009calculations}
Marrel, A., Iooss, B., Laurent, B., and Roustant, O. (2009).
\newblock Calculations of {S}obol indices for the {G}aussian process metamodel.
\newblock {\em Reliability Engineering \& System Safety}, 94(3):742--751.

\bibitem[McCulloch et~al., 2019]{BART}
McCulloch, R., Sparapani, R., Gramacy, R., Spanbauer, C., and Pratola, M.
  (2019).
\newblock {\em BART: {B}ayesian Additive Regression Trees}.
\newblock R package version 2.7.

\bibitem[McKay et~al., 1979]{mckay1979comparison}
McKay, M.~D., Beckman, R.~J., and Conover, W.~J. (1979).
\newblock Comparison of three methods for selecting values of input variables
  in the analysis of output from a computer code.
\newblock {\em Technometrics}, 21(2):239--245.

\bibitem[McKeague et~al., 2005]{mckeague2005statistical}
McKeague, I.~W., Nicholls, G., Speer, K., and Herbei, R. (2005).
\newblock Statistical inversion of south atlantic circulation in an abyssal
  neutral density layer.
\newblock {\em Journal of Marine Research}, 63(4):683--704.

\bibitem[McKinley et~al., 2018]{Mckinley2018approximate}
McKinley, T.~J., Vernon, I., Andrianakis, I., McCreesh, N., Oakley, J.~E.,
  Nsubuga, R.~N., Goldstein, M., White, R.~G., et~al. (2018).
\newblock Approximate {B}ayesian computation and simulation-based inference for
  complex stochastic epidemic models.
\newblock {\em Statistical Science}, 33(1):4--18.

\bibitem[Mockus et~al., 1978]{Mockus1978}
Mockus, J., Tiesis, V., and Zilinskas, A. (1978).
\newblock The application of {B}ayesian methods for seeking the extremum.
\newblock {\em Towards Global Optimization}, 2(117-129):2.

\bibitem[Molina et~al., 2005]{molina2005statistical}
Molina, G., Bayarri, M.~J., and Berger, J.~O. (2005).
\newblock Statistical inverse analysis for a network microsimulator.
\newblock {\em Technometrics}, 47(4):388--398.

\bibitem[Mortveit et~al., 2015]{mortveit2015synthetic}
Mortveit, H., Adiga, A., Agashe, A., Alam, M., Alexander, K., Arifuzzaman, S.,
  Barrett, C., Beckman, R., Bisset, K., Chen, J., et~al. (2015).
\newblock Synthetic populations and interaction networks for guinea. liberia
  and sierra leone.
\newblock Technical report, NDSSL.

\bibitem[Moutoussamy et~al., 2015]{Moutoussamy2015}
Moutoussamy, V., Nanty, S., and Pauwels, B. (2015).
\newblock Emulators for stochastic simulation codes.
\newblock {\em ESAIM: Proceedings and Surveys}, 48:116--155.

\bibitem[Neal, 1996]{neal_bayesian_1996}
Neal, R.~M. (1996).
\newblock {\em {B}ayesian {Learning} for {Neural} {Networks}}, volume 118 of
  {\em Lecture {Notes} in {Statistics}}.
\newblock Springer, New York, New York, NY.

\bibitem[Nguyen et~al., 2017]{nguyen2017multivariate}
Nguyen, H., Cressie, N., and Braverman, A. (2017).
\newblock Multivariate spatial data fusion for very large remote sensing
  datasets.
\newblock {\em Remote Sensing}, 9(2):142.

\bibitem[Oakley and O'Hagan, 2004]{oakley2004probabilistic}
Oakley, J.~E. and O'Hagan, A. (2004).
\newblock Probabilistic sensitivity analysis of complex models: a {B}ayesian
  approach.
\newblock {\em Journal of the Royal Statistical Society: Series B (Statistical
  Methodology)}, 66(3):751--769.

\bibitem[Oakley and Youngman, 2017]{Oakley2017calibration}
Oakley, J.~E. and Youngman, B.~D. (2017).
\newblock Calibration of stochastic computer simulators using likelihood
  emulation.
\newblock {\em Technometrics}, 59(1):80--92.

\bibitem[O'Hagan, 2006]{OHagan2006}
O'Hagan, A. (2006).
\newblock {B}ayesian analysis of computer code outputs: A tutorial.
\newblock {\em Reliability Engineering and System Safety},
  91(10-11):1290--1300.

\bibitem[Opitz et~al., 2018]{opitz2018inla}
Opitz, T., Huser, R., Bakka, H., and Rue, H. (2018).
\newblock {INLA} goes extreme: {B}ayesian tail regression for the estimation of
  high spatio-temporal quantiles.
\newblock {\em Extremes}, 21(3):441--462.

\bibitem[Papamakarios et~al., 2019]{papamakarios2019normalizing}
Papamakarios, G., Nalisnick, E., Rezende, D.~J., Mohamed, S., and
  Lakshminarayanan, B. (2019).
\newblock Normalizing flows for probabilistic modeling and inference.
\newblock {\em arXiv preprint arXiv:1912.02762}.

\bibitem[Park and Apley, 2018]{park2018patchwork}
Park, C. and Apley, D. (2018).
\newblock Patchwork kriging for large-scale {G}aussian process regression.
\newblock {\em The Journal of Machine Learning Research}, 19(1):269--311.

\bibitem[Paulo et~al., 2012]{paulo2012calibration}
Paulo, R., Garc{\'\i}a-Donato, G., and Palomo, J. (2012).
\newblock Calibration of computer models with multivariate output.
\newblock {\em Computational Statistics \& Data Analysis}, 56(12):3959--3974.

\bibitem[Pedregosa et~al., 2011]{scikit-learn}
Pedregosa, F., Varoquaux, G., Gramfort, A., Michel, V., Thirion, B., Grisel,
  O., Blondel, M., Prettenhofer, P., Weiss, R., Dubourg, V., Vanderplas, J.,
  Passos, A., Cournapeau, D., Brucher, M., Perrot, M., and Duchesnay, E.
  (2011).
\newblock Scikit-learn: Machine learning in {P}ython.
\newblock {\em Journal of Machine Learning Research}, 12:2825--2830.

\bibitem[Peleg et~al., 2017]{peleg2017advanced}
Peleg, N., Fatichi, S., Paschalis, A., Molnar, P., and Burlando, P. (2017).
\newblock An advanced stochastic weather generator for simulating 2-d
  high-resolution climate variables.
\newblock {\em Journal of Advances in Modeling Earth Systems}, 9(3):1595--1627.

\bibitem[Picheny and Ginsbourger, 2014]{picheny2014noisy}
Picheny, V. and Ginsbourger, D. (2014).
\newblock Noisy kriging-based optimization methods: a unified implementation
  within the diceoptim package.
\newblock {\em Computational Statistics \& Data Analysis}, 71:1035--1053.

\bibitem[Picheny et~al., 2016]{DiceOptim}
Picheny, V., Ginsbourger, D., and Roustant, O. (2016).
\newblock {\em DiceOptim: Kriging-Based Optimization for Computer Experiments}.
\newblock R package version 2.0.

\bibitem[Picheny et~al., 2013]{picheny2013benchmark}
Picheny, V., Wagner, T., and Ginsbourger, D. (2013).
\newblock A benchmark of kriging-based infill criteria for noisy optimization.
\newblock {\em Structural and Multidisciplinary Optimization}, 48(3):607--626.

\bibitem[Plumlee, 2017]{plumlee2017bayesian}
Plumlee, M. (2017).
\newblock {B}ayesian calibration of inexact computer models.
\newblock {\em Journal of the American Statistical Association},
  112(519):1274--1285.

\bibitem[Plumlee and Tuo, 2014]{Plumlee2014building}
Plumlee, M. and Tuo, R. (2014).
\newblock Building accurate emulators for stochastic simulations via quantile
  kriging.
\newblock {\em Technometrics}, 56(4):466--473.

\bibitem[Poole and Raftery, 2000]{poole2000inference}
Poole, D. and Raftery, A.~E. (2000).
\newblock Inference for deterministic simulation models: the {B}ayesian melding
  approach.
\newblock {\em Journal of the American Statistical Association},
  95(452):1244--1255.

\bibitem[Pratola et~al., 2020]{pratola2020heteroscedastic}
Pratola, M., Chipman, H., George, E., and McCulloch, R. (2020).
\newblock Heteroscedastic {BART} via multiplicative regression trees.
\newblock {\em Journal of Computational and Graphical Statistics},
  29(2):405--417.

\bibitem[Pronzato and M{\"u}ller, 2011]{Pronzato2011}
Pronzato, L. and M{\"u}ller, W.~G. (2011).
\newblock {Design of computer experiments: space filling and beyond}.
\newblock {\em Statistics and Computing}, 22(3):681--701.

\bibitem[Pukelsheim, 1994]{pukelsheim1994three}
Pukelsheim, F. (1994).
\newblock The three sigma rule.
\newblock {\em The American Statistician}, 48(2):88--91.

\bibitem[Qian et~al., 2008]{qian2008gaussian}
Qian, P. Z.~G., Wu, H., and Wu, C.~J. (2008).
\newblock {G}aussian process models for computer experiments with qualitative
  and quantitative factors.
\newblock {\em Technometrics}, 50(3):383--396.

\bibitem[Radtke et~al., 2002]{radtke2002bayesian}
Radtke, P.~J., Burk, T.~E., and Bolstad, P.~V. (2002).
\newblock {B}ayesian melding of a forest ecosystem model with correlated
  inputs.
\newblock {\em Forest Science}, 48(4):701--711.

\bibitem[Raftery et~al., 1995]{raftery1995inference}
Raftery, A.~E., Givens, G.~H., and Zeh, J.~E. (1995).
\newblock Inference from a deterministic population dynamics model for bowhead
  whales.
\newblock {\em Journal of the American Statistical Association},
  90(430):402--416.

\bibitem[Ramsey and Efford, 2010]{ramsey2010management}
Ramsey, D.~S. and Efford, M.~G. (2010).
\newblock Management of bovine tuberculosis in brushtail possums in new
  zealand: predictions from a spatially explicit, individual-based model.
\newblock {\em Journal of Applied Ecology}, 47(4):911--919.

\bibitem[Rannou et~al., 2002]{rannou2002kriging}
Rannou, V., Brouaye, F., H{\'e}lier, M., and Tabbara, W. (2002).
\newblock Kriging the quantile: application to a simple transmission line
  model.
\newblock {\em Inverse Problems}, 18(1):37.

\bibitem[Rasmussen and Williams, 2006]{rasmussen2006gaussian}
Rasmussen, C.~E. and Williams, C.~K. (2006).
\newblock {\em {G}aussian Processes for Machine Learning}.
\newblock MIT press, Cambridge, MA.

\bibitem[Reynolds, 1987]{reynolds1987flocks}
Reynolds, C.~W. (1987).
\newblock Flocks, herds and schools: A distributed behavioral model.
\newblock In {\em Proceedings of the 14th Annual Conference on Computer
  Graphics and Interactive Techniques}, pages 25--34.

\bibitem[Richardson, 1981]{richardson1981stochastic}
Richardson, C.~W. (1981).
\newblock Stochastic simulation of daily precipitation, temperature, and solar
  radiation.
\newblock {\em Water Resources Research}, 17(1):182--190.

\bibitem[Roustant et~al., 2018]{DiceKriging}
Roustant, O., Ginsbourger, D., and Deville., Y. (2018).
\newblock {\em {\tt {DiceKriging}: Kriging Methods for Computer Experiments}}.
\newblock R package version 1.5.6.

\bibitem[Rulli{\`e}re et~al., 2018]{rulliere2018nested}
Rulli{\`e}re, D., Durrande, N., Bachoc, F., and Chevalier, C. (2018).
\newblock Nested kriging predictions for datasets with a large number of
  observations.
\newblock {\em Statistics and Computing}, 28(4):849--867.

\bibitem[Rushdi et~al., 2017]{rushdi2017vps}
Rushdi, A., Swiler, L., Phipps, E., D'Elia, M., and Ebeida, M. (2017).
\newblock {VPS}: {V}oronoi piecewise surrogate models for high-dimensional data
  fitting.
\newblock {\em International Journal for Uncertainty Quantification}, 7(1).

\bibitem[Rutter et~al., 2019]{rutter2018microsimulation}
Rutter, C.~M., Ozik, J., DeYoreo, M., Collier, N., et~al. (2019).
\newblock Microsimulation model calibration using incremental mixture
  approximate {B}ayesian computation.
\newblock {\em The Annals of Applied Statistics}, 13(4):2189--2212.

\bibitem[Sacks et~al., 1989]{sacks1989design}
Sacks, J., Welch, W.~J., Mitchell, T.~J., and Wynn, H.~P. (1989).
\newblock Design and analysis of computer experiments.
\newblock {\em Statistical Science}, 4(4):409--423.

\bibitem[Salter et~al., 2019]{salter2019uncertainty}
Salter, J.~M., Williamson, D.~B., Scinocca, J., Kharin, V., et~al. (2019).
\newblock Uncertainty quantification for computer models with spatial output
  using calibration-optimal bases.
\newblock {\em Journal of the American Statistical Association},
  114(528):1800--1814.

\bibitem[Santner et~al., 2018]{santner2018design}
Santner, T.~J., B., W., and W., N. (2018).
\newblock {\em The Design and Analysis of Computer Experiments, Second
  Edition}.
\newblock Springer.

\bibitem[Schonlau and Welch, 2006]{schonlau2006screening}
Schonlau, M. and Welch, W.~J. (2006).
\newblock Screening the input variables to a computer model via analysis of
  variance and visualization.
\newblock In {\em Screening}, pages 308--327. Springer.

\bibitem[Schultz and Sokolov, 2018]{schultz2018practical}
Schultz, L. and Sokolov, V. (2018).
\newblock Practical {B}ayesian optimization for transportation simulators.
\newblock {\em arXiv preprint arXiv:1810.03688}.

\bibitem[{\v{S}}ev{\v{c}}{\'\i}kov{\'a} et~al., 2007]{vsevvcikova2007assessing}
{\v{S}}ev{\v{c}}{\'\i}kov{\'a}, H., Raftery, A.~E., and Waddell, P.~A. (2007).
\newblock Assessing uncertainty in urban simulations using {B}ayesian melding.
\newblock {\em Transportation Research Part B: Methodological}, 41(6):652--669.

\bibitem[Shah et~al., 2014]{shah2014student}
Shah, A., Wilson, A., and Ghahramani, Z. (2014).
\newblock Student-t processes as alternatives to {G}aussian processes.
\newblock In {\em Proceedings of the 17th International Conference on
  Artificial Intelligence and Statistics}, volume~33 of {\em PMLR}, pages
  877--885.

\bibitem[Smieszek et~al., 2011]{smieszek2011reconstructing}
Smieszek, T., Balmer, M., Hattendorf, J., Axhausen, K.~W., Zinsstag, J., and
  Scholz, R.~W. (2011).
\newblock Reconstructing the 2003/2004 h3n2 influenza epidemic in switzerland
  with a spatially explicit, individual-based model.
\newblock {\em BMC Nnfectious Diseases}, 11(1):115.

\bibitem[Sobol, 1967]{sobol1967distribution}
Sobol, I.~M. (1967).
\newblock On the distribution of points in a cube and the approximate
  evaluation of integrals.
\newblock {\em Computational Mathematics and Mathematical Physics},
  7(4):86--112.

\bibitem[Sobol, 1993]{sobol1993sensitivity}
Sobol, I.~M. (1993).
\newblock Sensitivity estimates for nonlinear mathematical models.
\newblock {\em Mathematical Modelling and Computational Experiments},
  1(4):407--414.

\bibitem[Spiller et~al., 2014]{spiller2014automating}
Spiller, E.~T., Bayarri, M., Berger, J.~O., Calder, E.~S., Patra, A.~K.,
  Pitman, E.~B., and Wolpert, R.~L. (2014).
\newblock Automating emulator construction for geophysical hazard maps.
\newblock {\em SIAM/ASA Journal on Uncertainty Quantification}, 2(1):126--152.

\bibitem[Srinivas et~al., 2009]{srinivas2009gaussian}
Srinivas, N., Krause, A., Kakade, S.~M., and Seeger, M. (2009).
\newblock {G}aussian process optimization in the bandit setting: No regret and
  experimental design.
\newblock {\em arXiv preprint arXiv:0912.3995}.

\bibitem[Stein, 2012]{stein2012interpolation}
Stein, M.~L. (2012).
\newblock {\em Interpolation of Spatial data: Some Theory for Kriging}.
\newblock Springer Science \& Business Media.

\bibitem[Sullivan, 2015]{sullivan2015introduction}
Sullivan, T.~J. (2015).
\newblock {\em Introduction to Uncertainty Quantification}, volume~63.
\newblock Springer.

\bibitem[Sun et~al., 2019]{sun2019synthesizing}
Sun, F., Gramacy, R., Haaland, B., Lu, S., and Hwang, Y. (2019).
\newblock Synthesizing simulation and field data of solar irradiance.
\newblock {\em Statistical Analysis and Data Mining}, 12(4):311--324.
\newblock Preprint on ar{X}iv:1806.05131.

\bibitem[Sung et~al., 2019]{sung2019calibration}
Sung, C.-L., Barber, B.~D., and Walker, B.~J. (2019).
\newblock Calibration of computer models with heteroscedastic errors and
  application to plant relative growth rates.
\newblock {\em arXiv preprint arXiv:1910.11518}.

\bibitem[Tuo and Wu, 2016]{tuo2016theoretical}
Tuo, R. and Wu, C. (2016).
\newblock A theoretical framework for calibration in computer models:
  parametrization, estimation and convergence properties.
\newblock {\em SIAM/ASA Journal on Uncertainty Quantification}, 4(1):767--795.

\bibitem[Tuo et~al., 2015]{tuo2015efficient}
Tuo, R., Wu, C.~J., et~al. (2015).
\newblock Efficient calibration for imperfect computer models.
\newblock {\em The Annals of Statistics}, 43(6):2331--2352.

\bibitem[Vazquez et~al., 2008]{Vazquez2008}
Vazquez, E., Villemonteix, J., Sidorkiewicz, M., and Walter, E. (2008).
\newblock Global optimization based on noisy evaluations: an empirical study of
  two statistical approaches.
\newblock In {\em Journal of Physics: Conference Series}, volume 135, page
  012100. IOP Publishing.

\bibitem[Vernon et~al., 2010]{vernon2010galaxy}
Vernon, I., Goldstein, M., Bower, R.~G., et~al. (2010).
\newblock Galaxy formation: a {B}ayesian uncertainty analysis.
\newblock {\em {B}ayesian Analysis}, 5(4):619--669.

\bibitem[Wang and Ng, 2020]{wang2020enhancing}
Wang, S. and Ng, S.~h. (2020).
\newblock Enhancing response predictions with a joint {G}aussian process model
  for stochastic simulation models.
\newblock {\em ACM Transactions on Modeling and Computer Simulation (TOMACS)},
  30(1):1--25.

\bibitem[Wang and Haaland, 2019]{wang2019controlling}
Wang, W. and Haaland, B. (2019).
\newblock Controlling sources of inaccuracy in stochastic kriging.
\newblock {\em Technometrics}, 61(3):309--321.

\bibitem[Wang et~al., 2017]{wang2017extended}
Wang, Z., Shi, J.~Q., and Lee, Y. (2017).
\newblock Extended t-process regression models.
\newblock {\em Journal of Statistical Planning and Inference}, 189:38--60.

\bibitem[Welling and Teh, 2011]{welling2011bayesian}
Welling, M. and Teh, Y.~W. (2011).
\newblock {B}ayesian learning via stochastic gradient {L}angevin dynamics.
\newblock In {\em Proceedings of the 28th International Conference on Machine
  Learning (ICML-11)}, pages 681--688.

\bibitem[Wilensky, 1999]{Wilensky1999}
Wilensky, U. (1999).
\newblock {NetLogo. http://ccl.northwestern.edu/netlogo/.}
\newblock {\em Center for Connected Learning and ComputerBased Modeling
  Northwestern University Evanston IL}.

\bibitem[Wilkinson, 2013]{wilkinson2013approximate}
Wilkinson, R.~D. (2013).
\newblock Approximate {B}ayesian computation (abc) gives exact results under
  the assumption of model error.
\newblock {\em Statistical Applications in Genetics and Molecular Biology},
  12(2):129--141.

\bibitem[Xie and Chen, 2017]{Xie2017}
Xie, G. and Chen, X. (2017).
\newblock A heteroscedastic t-process simulation metamodeling approach and its
  application in inventory control and optimization.
\newblock In {\em Simulation Conference (WSC), 2017 Winter}, pages 3242--3253.
  IEEE.

\bibitem[Yohan~Chalabi and Wuertz, 2019]{randtoolbox}
Yohan~Chalabi, Christophe~Dutang, P.~S. and Wuertz, D. (2019).
\newblock {\em randtoolbox: Toolbox for Pseudo and Quasi Random Number
  Generation and Random Generator Tests}.
\newblock R package version 1.30.0.

\bibitem[Zhang et~al., 2020]{zhang2018distance}
Zhang, B., Cole, D., and Gramacy, R. (2020).
\newblock Distance-distributed design for {G}aussian process surrogates.
\newblock {\em To appear in Technometrics}.
\newblock Preprint on ar{X}iv:1812.02794.

\bibitem[Zhang et~al., 2008]{zhang2008loss}
Zhang, J., Craigmile, P.~F., and Cressie, N. (2008).
\newblock Loss function approaches to predict a spatial quantile and its
  exceedance region.
\newblock {\em Technometrics}, 50(2):216--227.

\bibitem[Zhang and Xie, 2017]{zhang2017asymmetric}
Zhang, Q. and Xie, W. (2017).
\newblock Asymmetric kriging emulator for stochastic simulation.
\newblock In {\em Proceedings of the 2017 Winter Simulation Conference}, page
  137. IEEE Press.

\bibitem[Zhang et~al., 2018]{Zhang2019latent}
Zhang, Y., Tao, S., Chen, W., and Apley, D.~W. (2018).
\newblock A latent variable approach to {G}aussian process modeling with
  qualitative and quantitative factors.
\newblock {\em arXiv preprint arXiv:1806.07504}.

\bibitem[Zhou et~al., 2012]{zhou2012estimating}
Zhou, J., Chang, H.~H., and Fuentes, M. (2012).
\newblock Estimating the health impact of climate change with calibrated
  climate model output.
\newblock {\em Journal of Agricultural, Biological, and Environmental
  Statistics}, 17(3):377--394.

\bibitem[Zhou et~al., 2011]{zhou2011simple}
Zhou, Q., Qian, P.~Z., and Zhou, S. (2011).
\newblock A simple approach to emulation for computer models with qualitative
  and quantitative factors.
\newblock {\em Technometrics}, 53(3):266--273.

\end{thebibliography}

\end{document}